\documentclass{jfm}
\usepackage{graphicx}
\usepackage{epstopdf, epsfig}
\usepackage{comment} 
\usepackage{amsmath} 
\usepackage{color}

\shorttitle{Negative APE dissipation and double diffusive instabilities}
\shortauthor{R. Tailleux}

\title{
Negative APE dissipation as the fundamental criterion for double diffusive instabilities}

\author{R. Tailleux\aff{1}
  \corresp{\email{R.G.J.Tailleux@reading.ac.uk}},
}

\affiliation{\aff{1}Department of Meteorology, University of Reading, Reading RG6 6BB, UK}

\begin{document}

\maketitle

\begin{abstract}
\textcolor{black}{
The background potential energy (BPE) is the only reservoir that double diffusive instabilities can tap their energy from when developing from an unforced motionless state with no available potential energy (APE). Recently, Middleton and Taylor linked the extraction of BPE into APE to the sign of the diapycnal component of the buoyancy flux, but their criterion can only predict diffusive convection instability, not salt finger instability. Here, we show that the problem can
be corrected if the sign of the APE dissipation rate is used instead, making it
emerge as the most fundamental criterion for double diffusive instabilities. A theory for the APE dissipation rate for a two-component fluid relative to its single-component counterpart is developed as a function of three parameters: the diffusivity ratio, the density ratio, and a spiciness parameter. The theory correctly predicts the occurrence of both salt finger and diffusive convection instabilities in the laminar unforced regime, 
while more generally predicting that the APE dissipation rate for a two-component fluid can be enhanced, suppressed, or even have the opposite sign compared to that for a single-component fluid, with important implications for the study of ocean mixing. Because negative APE dissipation can also occur 
in stably stratified 
single-component and doubly stable two-component stratified fluids, we speculate that only the thermodynamic
theory of exergy can explain its physics; however, this necessitates accepting that APE dissipation is a conversion between APE and the internal energy component of BPE, in contrast to prevailing assumptions.  
}

\end{abstract}


\section{Introduction}

Stratified binary fluids, in which the stratifying agents diffuse at different rates, exhibit a range of intriguing phenomena not observed in single-component stratified fluids. One such phenomenon are double diffusive instabilities, which arise when one of the stratifying agents act to destabilise the stratification. In the context of ocean mixing, this leads to the emergence of new types of flows, including diffusive convection, salt fingering, thermohaline intrusions, and thermohaline staircases (Turner, 1985; Schmitt, 1994; Radko, 2013). Theoretical analysis categorises the different dynamically relevant types of stratification based on the density ratio $(R_{\rho})$ of the fluid: 1) the doubly stable case $(R_{\rho}<0)$, when both temperature and salinity are stabilising; 2) salt fingers favourable $(R_{\rho}>1)$, with salinity as the destabilising agent; 3) diffusive convection favourable $(0<R_{\rho}<1)$, where temperature acts as the destabiliser. Understanding how double diffusion modulates mixing in the doubly stable case and characterising double diffusive instabilities in both laminar and turbulent regimes are key questions of interest.  

To date, linear stability analysis has been the primary approach for identifying the physical parameters that control the development of double diffusive instabilities from a quiescent sate. Such an analysis reveals, for instance, that molecular viscosity and diffusion can severely restrict the range of density ratios at which these instabilities can occur, viz.,   
\begin{equation}
\begin{split}
    \frac{Pr+ \tau}{Pr + 1} < R_{\rho} < 1, & \qquad 
    {\rm for\,\, diffusive \,\, convection} ,\\
    1 < R_{\rho} < \frac{1}{\tau}, & \qquad 
    {\rm for \,\, salt \,\,fingers} ,
\end{split}
\label{dd_regimes} 
\end{equation}
e.g., \citet{Radko2013}, where $\tau = \kappa_S/\kappa_T$ represents the ratio of salt diffusivity to temperature diffusivity, $Pr = \nu/\kappa_T$ denotes the Prandtl number, and $\nu$ is the molecular viscosity. For example, using the typical oceanic values $\tau \approx 0.01$ and $Pr \approx 7$ yield a reduced density ratio range of $1 < R_{\rho} < 100$ for salt fingers and $0.8076 < R_{\rho} < 1$ for diffusive convection.  

While linear stability theory provides valuable insights into double diffusive instabilities, other approaches are necessary to understand their energy source and whether they can persist in a turbulent environment like the oceans. In this regard, the theory of available potential energy (APE) introduced by \citet{Lorenz1955} and applied to the study of turbulent stratified fluids by \citet{Winters1995} (referred to as W95 thereafter) appears to be one of the most promising avenues of research. According to APE theory, the energetics of a stratified fluid can be separated into two main budgets, one for the total mechanical energy (APE+kinetic energy), and one for the background potential energy (BPE). Physically, the BPE represents the potential energy (PE) associated with the state of minimum potential energy achievable from the actual state through an adiabatic (and isohaline in seawater) mass re-arrangement. The generic form of these budget equations is
\begin{equation}
       \frac{d(APE+KE)}{dt} = {\rm Forcing}
       - \int_{V} \rho (\varepsilon_k + \varepsilon_p )\,{\rm d}V
\end{equation}
\begin{equation}
       \frac{d(BPE)}{dt} = \int_{V} \rho (\varepsilon_k + \varepsilon_p ) \,{\rm d}V 
\end{equation}
where $\varepsilon_k$ is the positive definite viscous dissipation rate, and $\varepsilon_p$ the diffusive rate of APE dissipation. As is well known, viscous dissipation is an irreversible process that converts KE into `heat', which in APE theory is played by the BPE. 

In a single-component compressible fluid, \citet{Tailleux2013c} 
\textcolor{black}{(of which an improved version is presented in Appendix A)} 
established that the APE dissipation rate $\varepsilon_p$ may be broken down into three components:
\begin{equation}
     \varepsilon_p = \varepsilon_{p,lam} + \varepsilon_{p,tur} + \varepsilon_{p,eos} ,
\end{equation}
with $\varepsilon_{p,lam}$ and $\varepsilon_{p,tur}$ representing the laminar and turbulent parts of $\varepsilon_p$ respectively, and $\varepsilon_{p,eos}$ representing the part of $\varepsilon_p$ arising from the nonlinearities of the equation of state. 
To date, the APE dissipation rate has been primarily defined and studied for single-component Boussinesq fluids with a linear equation of state, for wich only $\varepsilon_{p,lam}$ and $\varepsilon_{p,tur}$ subsist. As clarified further in the text, $\varepsilon_{p,lam}$ and $\varepsilon_{p,tur}$ relate to the terms $-\Phi_i$ and $\Phi_d$ in W95's framework. Physically, $\varepsilon_{p,tur}$ is the fundamental quantity for defining the turbulent diapycnal diffusivity 
\begin{equation}
        K_{\rho} = \frac{\varepsilon_{p,tur}}{N^2} = \frac{\Gamma \varepsilon_k}{N^2} 
\end{equation}
e.g., \citet{Osborn1972,Osborn1980,Lindborg2008}, often quantified by the dissipation ratio or `mixing efficiency' $\Gamma = \varepsilon_{p,tur}/\varepsilon_k$. Typically, $\varepsilon_{p,tur}$ is estimated from microstructure measurements using formula such as the widely used expression: 
\begin{equation}
        \varepsilon_{p,tur}
        =  \frac{g \alpha \kappa_T |\nabla \theta'|^2}{d\overline{\theta}/dz} ,
        \label{APE_dissipation_measured} 
\end{equation}
\citep{Oakey1982,Gargett1984b}, where $\alpha$ is the thermal expansion coefficient, $g$ is the acceleration due to gravity, and $\overline{\theta}$ represents the mean potential temperature profile. Consequently, $\varepsilon_p$, like the viscous dissipation rate, is generally considered a positive quantity that also converts mechanical energy into BPE or `heat' (for further views on this, see \citet{Tailleux2009}).  

In a double diffusive fluid, double diffusive instabilities can develop even in the absence of mechanical forcing. Within the framework of APE theory, this is only possible if BPE, like APE, can become a source of kinetic energy, and hence if we accept the notion \textcolor{black}{that $\varepsilon_p$ can become negative}. In this paper, we derive a theoretical expression for the ratio ${\cal F}= \varepsilon_{p}^{dd}/\varepsilon_p^{std}$ of the APE dissipation rates for a double diffusive case over that for a simple fluid.
We demonstrate that $\varepsilon_p^{dd}$ is negative under the conditions corresponding to the linear stability analysis for both salt finger and diffusive convection regimes. Recently, \citet{Middleton2020} proposed to characterise double diffusive instabilities in terms of the sign of the diapycnal component of the molecular buoyancy flux, which is related to the turbulent part $\varepsilon_{p,tur}$ of $\varepsilon_p$ (equivalently $\Phi_d$), rather than $\varepsilon_p$. 
However, their criterion only succeeds in predicting the occurrence of the diffusive convection instability but not the salt finger instability, which suggests that it is the sign of the net APE dissipation rate $\varepsilon_{p,lam} + \varepsilon_{p,tur}$ (equivalently $\Phi_d-\Phi_i$) that represents the most fundamental criterion for double diffusive instabilities. Additionally, we show that considering the APE dissipation rate sheds light on other aspects of double diffusive instabilities in turbulent regimes. For example, it correctly predicts the regime associated with diffusive interleaving in the presence of density-compensated isopycnal gradients of temperature and salinity, which are thought to be essential for the formation of thermohaline staircases \citep{Merryfield2000}. 

The remainder of this paper is organised as follows: Section 2 establishes the general form of the local APE dissipation rate for a double diffusive fluid. Section 3 outlines the conditions necessary for negative APE dissipation. Section 4 provides a comparison of the local and APE frameworks. Section 5 discusses the practical issues related to the computation of the reference temperature and salinity profiles used in our definition of spiciness. Finally, Section 6 discusses the results and provides future perspectives.

\section{Local APE theory and APE dissipation rate}
\label{energetics_binary_fluids} 

Similarly as \citet{Middleton2020}, we study 
the energetics of double diffusive instabilities for a 
Boussinesq fluid with a linear equation of state for density governed by the system of equations:
\begin{equation}
     \frac{D{\bf v}}{Dt} + \frac{1}{\rho_{\star}} \nabla \delta p
     = - \frac{g \delta \rho}{\rho_{\star}} 
     {\bf k} +  \nu \nabla^2 {\bf v}, 
\end{equation}
\begin{equation}
      \nabla \cdot {\bf v} = 0, 
\end{equation}
\begin{equation}
       \frac{D\theta}{Dt} = - \nabla \cdot {\bf J}_{\theta}, \qquad
       \frac{DS}{Dt} = -\nabla \cdot {\bf J}_S ,
       \label{tracer_equations} 
\end{equation}
\begin{equation}
    \rho = \rho_{\star} \left [ 1 - \alpha (\theta - \theta_{\star} ) 
    + \beta (S-S_{\star} ) \right ] ,
    \label{eos} 
\end{equation}
where $\rho$ is density, $p$ is pressure, $g$ is gravitational acceleration, 
$S$ is salinity, $\theta$ is potential temperature, $\alpha$ is the thermal expansion coefficient, $\beta$ is the haline contraction coefficients, while $\rho_{\star}$, $S_{\star} $ and $\theta_{\star}$ are constant reference values for $\rho$, $S$ and $\theta$ respectively. 
Moreover, $\delta p = p-p_0(z)$ and $\delta \rho = \rho - \rho_0(z)$ denote the pressure and
density anomalies defined relative to the reference pressure and density profiles
$p_0(z)$ and $\rho_0(z) = -g^{-1} dp_0/dz$ characterising Lorenz reference state.
Simple diffusive laws are assumed for $\theta$ and $S$ so that ${\bf J}_{\theta} = -\kappa_T \nabla \theta$ and ${\bf J}_s = -\kappa_S \nabla S$, where $\kappa_T$ and $\kappa_S$ are the molecular diffusivities of heat and salt respectively. 
The tracer equations (\ref{tracer_equations}) 
may be combined with the equation of state (\ref{eos}) to form an equation
for the Boussinesq buoyancy $b_{bou} = g [ \alpha (\theta-\theta_{\star}) - \beta 
(S-S_{\star})]$, 
\begin{equation}
     \frac{Db_{bou}}{Dt} = - \nabla \cdot {\bf J}_b 
\end{equation}
where 
\begin{equation}
    {\bf J}_b = g (\alpha {\bf J}_{\theta} - \beta {\bf J}_s ) 
    = - g (\kappa_T \alpha \nabla \theta - \kappa_S \beta \nabla S)
\end{equation}
is the molecular buoyancy flux. Note that in the absence of salinity, ${\bf J}_b = 
-\kappa_T \nabla b_{bou}$ is down the gradient of the Boussinesq buoyancy, but not
in the general case.

To understand how double diffusion affects APE dissipation, the local APE framework \citep{Tailleux2013b,Tailleux2018,Novak2018} is preferred over the \citet{Winters1995}'s global APE framework used by \citet{Middleton2020}. Indeed, we show in Section \ref{interpretation} that defining the total APE of a fluid 
as the volume integral of non-sign definite
integrand makes the global APE framework problematic in several respects; for instance, 
it introduces unphysical terms in its budget. To avoid such difficulties, it is 
therefore preferable to define the APE of a fluid as the volume integral of the APE density 
\begin{equation}
   e_a = e_a(S,\theta,z) = - \int_{z_r}^z b(S,\theta,z')\,{\rm d}z',
   \label{Boussinesq_ape_density}
\end{equation} 
where $b(S,\theta,z) = -(g/\rho_{\star}) (\rho(S,\theta)-\rho_0(z))$ is the buoyancy defined relative to Lorenz reference density profile $\rho_0(z)$ characterising the notional reference state of minimum potential energy obtainable from the actual state by means of an adiabatic and volume-conserving re-arrangement. This definition of buoyancy 
has the advantage of vanishing in Lorenz reference state of rest, which is not the case of
the Boussinesq buoyancy. \textcolor{black}{To avoid any ambiguity, note that 
$b(S,\theta,z') = -(g/\rho_{\star})( \rho(S(x,y,z,t),\theta(x,y,z,t))
- \rho_0(z'))$ in the integral (\ref{Boussinesq_ape_density}), that is, the values of $S$ and $\theta$ are held constant as the parcel moves from $z_r$ to $z$, consistent with the idea of an adiabatic and isohaline re-arrangement.} 
As is well known, the APE density (\ref{Boussinesq_ape_density})
represents the work against buoyancy forces that a parcel needs to counteract to move from its reference position at $z_r$,
solution of the level of neutral buoyancy equation
\begin{equation}
       b(S,\theta,z_r) = 0 ,
       \label{LNB_equation} 
\end{equation}
to its actual position at $z$. 
For simplicity, we disregard the temporal dependence of $\rho_0(z)$ as the APE dissipation only depends on spatial gradients of the tracer fields. Inversion of Eq. (\ref{LNB_equation}) establishes that $z_r= z_r(\rho) = z_r(S,\theta)$ is a material function of $S$ and $\theta$ and therefore a Lagrangian invariant following fluid parcels in the absence of diffusive sources/sinks of heat and salt.

A local budget equation for the APE density is easily obtained by taking the
material derivative of (\ref{Boussinesq_ape_density}), which yields
\begin{equation}
\begin{split} 
     \frac{De_a}{Dt} = & - b \frac{Dz}{Dt} + \underbrace{b(S,\theta,z_r)}_{=0} \frac{Dz_r}{Dt}
     - (z-z_r) \frac{Db_{ou}}{Dt}  \\ 
    = &  - b w - \nabla \cdot {\bf J}_a - \varepsilon_p , 
      \label{ape_density_evolution}
\end{split} 
\end{equation} 
where ${\bf J}_a$ is the diffusive flux of APE and $\varepsilon_p$ the APE dissipation rate,
given by
\begin{equation} 
    {\bf J}_a = -(z-z_r) {\bf J}_b ,
    \label{ape_density_diffusive_flux}
\end{equation}
\begin{equation} 
\begin{split} 
    \varepsilon_p = & {\bf J}_b \cdot 
        \nabla (z-z_r) \\ 
            = & \underbrace{- g \left ( \alpha \kappa_T \frac{\partial \theta}{\partial z} - \beta \kappa_S \frac{\partial S}{\partial z} \right )}_{\varepsilon_{p,lam}}  + 
    \underbrace{g (\alpha \kappa_T \nabla \theta - \beta \kappa_S \nabla S ) 
    \cdot \nabla z_r}_{\varepsilon_{p,tur}}  .  
        \label{APE_dissipation} 
\end{split} 
\end{equation}
As indicated by (\ref{APE_dissipation}), $\varepsilon_p$ can be decomposed into 
a `laminar' and `turbulent' components $\varepsilon_{p,lam}$ and $\varepsilon_{p,tur}$
whose volume integrals 
\begin{equation}
      \int_{V} \varepsilon_{p,lam} \rho_{\star} {\rm d}V = 
      - \Phi_i, \qquad 
      \int_{V} \varepsilon_{p,tur} \rho_{\star} {\rm d}V = \Phi_d ,
\end{equation}
can easily be shown to 
correspond to \citet{Winters1995} global energy conversions $-\Phi_i$ and $\Phi_d$ in 
the global APE framework. So far, the turbulent mixing community has generally assumed
$\Phi_i$ and $\Phi_d$ to represent 
distinct types of energy conversions, the former with internal energy (IE) and the latter with
the background potential energy (BPE); however, the analysis of
such conversions in real fluids detailed in Section \ref{interpretation}
clearly indicates that there is no physical basis for such a view, and that in reality, $\varepsilon_p$, $\Phi_d$, and $\Phi_i$ all represent conversions with internal energy 
as previously established by \citet{Tailleux2009}. In other words, $\varepsilon_{p,lam}$ and
$\varepsilon_{p,tur}$ are parts of the same energy conversion.

\smallskip 

Physically, Eq. (\ref{ape_density_evolution}) states that locally, $e_a$ can be modified through: 1) conversion with kinetic energy via the buoyancy flux $b w$; 2) diffusive and advective transport via ${\bf J}_a$ and advection; 3) `dissipation' that can be occasionally be negative associated with $\varepsilon_p$. Physically, our local APE budget (\ref{ape_density_evolution}) is 
simpler in form than the one previously derived by \citet{Scotti2014} for a single-component fluid,
although the two can be verified to be mathematically equivalent. One of the reasons
is due to \citet{Scotti2014} imposing the diffusive flux of APE density to be down-gradient, viz,
\begin{equation}
    {\bf J}_{a,sw14} = -\kappa \nabla e_a = \kappa (z-z_r) \nabla b_{bou} + \kappa b {\bf k} 
    = {\bf J}_a + \kappa b {\bf k} ,
    \label{downgradient_ape} 
\end{equation}
which differs from ${\bf J}_a$ by the last term, with the consequence of adding extra terms in their
budget equation. Physically, however, \citet{Scotti2014}'s approach is inconsistent with the 
analysis of the APE budget for a real fluid (cf Section \ref{interpretation} 
or \citet{Tailleux2009}), which reveals that decomposing
the potential energy into available and unavailable components leads to a decomposition of the heat
flux ${\bf J}_q = \Upsilon {\bf J}_q + (1-\Upsilon ) {\bf J}_q$, with $\Upsilon = (T-T_r)/T$ a
Carnot-like thermodynamic efficiency. Physically, it is the first term ${\bf }J_a = \Upsilon {\bf J}_q$ that represents the diffusive flux of APE density, whose Boussinesq approximation is given by
(\ref{ape_density_diffusive_flux}). Another reason to be sceptical of (\ref{downgradient_ape}) is 
that it cannot generalise to the case of a doubly-diffusive two-component fluid, 
whereas our definition of ${\bf J}_a$ (\ref{ape_density_diffusive_flux}) obviously does. 

\smallskip 

In their study, \citet{Middleton2020} derived an expression for the diapycnal
component of the buoyancy flux ${\bf J}_b\cdot \hat{\bf n}$ in terms of three parameters, 
the diffusivity ratio $\tau = \kappa_S/\kappa_T$, the gradient ratio $G_{\rho} = 
(\alpha |\nabla \theta|)/ (\beta |\nabla S)$, and the angle $\gamma$ 
between $\nabla \theta$ and $\nabla S$ such that $\cos{\gamma} = \nabla \theta \cdot \nabla S/(|\nabla \theta||\nabla S|)$, with $\hat{\bf n} = \nabla z_r/|\nabla z_r|$, and could similarly be used to
study $\varepsilon_p$.
However, while $G_{\rho}$ and $\cos{\gamma}$ might be easily diagnosed from the output of 
a numerical experiment, they cannot easily be 
inferred from measured oceanic properties nor are they naturally related to physical parameters 
known to play a key role in double diffusive instabilties such as density-compensated 
thermohaline variations. For instance, in the canonical salt finger experiment that they discuss in their Section 2.5, \citet{Middleton2020} find that 
$\Phi_d$ becomes negative as the result of the gradients of salinity growing more rapidly than temperature gradients, thus resulting in the creation of density-compensated thermohaline variations (usually referred to as `spice'). In oceanography, the term spice was perhaps first used by
\citet{Munk1981} to quantify the range of possible $\theta/S$ behaviour of a seawater sample of given density, from warm and salty (spicy) to fresh and cold (minty). Physically, the creation of spice is of considerable dynamical importance because lateral stirring along 
isopycnal surfaces is not opposed by restoring buoyancy forces and is therefore considerably more efficient at dissipating tracer variances than vertical stirring. To study the role of spice,
\citet{Middleton2021} subsequently reformulated \citet{Middleton2020}'s
expression in terms of the `spiciness'
variable $\tau = \alpha \theta + \beta S$ (denoted $S_p$ in their paper), which may be regarded
as the `linear' version of the spiciness variables developed by \citet{Jackett1985} or \citet{Flament2002} for instance. Recently, the theory of spiciness was revisited by 
\citet{Tailleux2021} who argued that to be physically meaningful, spice should ideally vanish in a spiceless ocean and hence that spice variables should be defined as isopycnal anomalies,
which is not the case of $\tau$ or of any related variable. In the 
present case, this can be implemented in practice by decomposing $\theta$ and $S$ as 
follows: 
\begin{equation}
        \theta = \theta_0 (z_r) + \theta_{\xi} , \qquad 
        S = S_0 (z_r) + S_{\xi} ,
\end{equation}
with the reference profiles $\theta_0(z_r)$ and $S_0(z_r)$ defining the temperature and salinity
profiles of the assumed spiceless state, with $z_r= z_r(\rho)$ denoting the reference position of a fluid parcel as before. Provided that $\theta_0(z_r)$ and $S_0(z_r)$ are defined as the \textcolor{black}{(thickness-weighted)} 
isopycnal mean temperature and salinity respectively, they contain all necessary information to compute density $\rho=\rho_{\star}[1-\alpha(\theta_0(z_r)-\theta_{\star}) - \beta (S_0(z_r)-S_{\star})]$, thus allowing them to be interpreted as the `active' components of $\theta$ and $S$. In contrast, the isopycnal anomalies $\theta_{\xi}$ and $S_{\xi}$ are density compensated $\alpha \theta_{\xi} = \beta S_{\xi}$ and therefore do not contribute to density, thus allowing them to be interpreted as the passive components of $\theta$ and $S$. Physically, defining the spiceless stratification $\theta_0(z),S_0(z)$ in terms of \textcolor{black}{thickness-weighted} isopycnal means can be justified on the ground that such a state
is the one that would be expected to result in the idealised limit of infinitely fast (resp. 
infinitely slow) isopycnal (resp. diapycnal) mixing. How to estimate these in practice are discussed in Section \ref{reference_state_variables}. As shown below, defining spiciness in this way allows
us to define a new set of parameters with which to study the behaviour of $\varepsilon_p$, different from those used by \citet{Middleton2020} and \citet{Middleton2021}, and which we consider to be 
more physical and more relevant. 

\smallskip 

In the following, $S_{\xi}$ and $\theta_{\xi}$ are combined into the variable $\xi = \rho_{\star} ( \alpha \theta_{\xi} + \beta S_{\xi}$) to define a single measure of spiciness having the dimension of density, as is commonly done in the literature. After some manipulation, it is possible to rewrite $\varepsilon_p$ in the following form
\smallskip
\par \noindent 
\fbox{
\begin{minipage}{13cm}
\begin{equation} 
\begin{split}
      \varepsilon_p = &
  g \left ( \kappa_T \alpha \frac{\partial \tilde{\theta}}{\partial z_r} 
  - \kappa_S \beta \frac{\partial \tilde{S}}{\partial z_r} \right ) 
  \left ( |\nabla z_r|^2 - \frac{\partial z_r}{\partial z} \right ) 
     + \frac{g ( \kappa_T - \kappa_S )}{2\rho_{\star}} \left ( \nabla z_r\cdot \nabla \xi - \frac{\partial \xi}{\partial z} \right ) \\
            = & \textcolor{black}{K_{v0}}
            N_0^2(z_r) \left [ \frac{R_{\rho} - \tau }{R_{\rho}-1} 
        + (1-\tau ) {\cal M} \right ]\left ( 1 - \frac{\partial_z z_r}{|\nabla z_r|^2} \right ) ,
        \label{epsilonp_final} 
\end{split}
\end{equation}
\end{minipage} 
}
\smallskip
\par \noindent 
which depends on the following parameters:
\par \medskip
\begin{equation}
\begin{split}
        {\cal M} \equiv \frac{g}{2\rho_{\star} N_0^2(z_r)}\frac{(\nabla z_r\cdot \nabla \xi -\partial_z \xi)}{|\nabla z_r|^2 -\partial_z z_r} \qquad & 
        {\rm Spiciness \,\,parameter} \\ \\ 
             \textcolor{black}{K_{v0}} \equiv \kappa_T |\nabla z_r|^2  
     = \kappa_T \left ( \frac{d \rho_0}{\partial z_r}(z_r) 
     \right )^{-2} |\nabla \rho|^2 \qquad & 
     {\rm Effective \,\, diffusivity} \\ \\
          N_0^2 (z_r) \equiv -\frac{g}{\rho_{\star}} \frac{d\rho_0}{dz_r}(z_r) = 
          g \left ( \alpha \frac{d\theta_0}{d z_r} 
      - \beta \frac{dS_0}{d z_r} \right ) (z_r) \qquad & 
      {\rm Reference \,\,N^2} \\ \\
            R_{\rho} \equiv R_{\rho}(z_r) = \left .  
      \alpha \frac{d\theta_0}{d z_r} \right/
      \beta \frac{dS_0}{d z_r} \qquad & 
      {\rm Density\,\,Ratio}   \\ \\ 
      \tau \equiv  \frac{\kappa_S}{\kappa_T}  \qquad & 
      {\rm Diffusivity\,\, Ratio} 
      \label{parameter_list} 
\end{split}
\end{equation}
Our choice of control parameters $R_{\rho}$ and ${\cal M}$ differ from those of \citet{Middleton2020}, who chose the gradient ratio $G_{\rho} \equiv \alpha |\nabla \theta |/(\beta |\nabla S|)$ and the angle $\phi$ between $\nabla \theta$ and $\nabla S$ defined by $\cos{\phi} = -\nabla \theta \cdot \nabla S/(|\nabla \theta | |\nabla S|)$ instead. The present approach is preferred here because $R_{\rho}$ is a more commonly encountered parameter in the literature, while ${\cal M}$ is more naturally connected to the physics of lateral thermohaline intrusions, whose importance is well established. Moreover, \textcolor{black}{$K_{v0}$} represents a local version of the effective diffusivity concept originally proposed by \citet{Winters1996} and \citet{Nakamura1996} \textcolor{black}{based on thermal diffusion}, and is positive definite by construction. 
\textcolor{black}{$K_{v0}$ is different from the effective diffusivity 
$K_{\rm eff}$ introduced in Section 5 and governing the evolution of the sorted
density profile.} 
Note that although the term within square brackets goes to infinity as $R_{\rho} \rightarrow 1$, $\varepsilon_p$ remains finite as $N_0^2 \rightarrow 0$ then.

\section{APE dissipation and diffusive stability} 
\label{stability_section} 

\subsection{Single component fluid} 

Even in the case of a single component fluid, the APE dissipation rate $\varepsilon_p$, unlike the viscous dissipation rate $\varepsilon_k$, is not guaranteed to be always positive definite. To see this, let us first set ${\cal M}=0$ and $\tau=1$ in (\ref{epsilonp_final}), thus leading to
\begin{equation}
       \varepsilon_p \equiv \varepsilon_p^{\rm std} = 
       \textcolor{black}{K_{v0}} N_0^2(z_r) 
       \left ( 1 - \frac{\partial_z z_r}{|\nabla z_r|^2} \right )  
       = \kappa_T \left ( |\nabla z_r|^2 - \frac{\partial z_r}{\partial z}
       \right ) N_0^2 .
       \label{epsilonp_single} 
\end{equation}
Since $\kappa_T$ and $N_0^2$ are positive by construction, (\ref{epsilonp_single}) shows that $\varepsilon_p$ can occasionally be negative
provided that 
\begin{equation}
        {\cal F} \equiv |\nabla z_r|^2 - \frac{\partial z_r}{\partial z} 
     = |\nabla_h z_r|^2 + 
     \left ( \frac{\partial z_r}{\partial z} \right )^2 
     - \frac{\partial z_r}{\partial z}  < 0 , 
\end{equation}
where $\nabla_h$ denotes the horizontal gradient. If viewed as a quadratic polynomial in $\partial z_r/\partial z$ with discriminant $\Delta = 1 - 4 |\nabla_h z_r|^2|$, ${\cal F}$ can only be negative when possessing two real roots, that is when $\Delta <0$, or equivalently when 
\begin{equation}
        |\nabla_h z_r| < \frac{1}{2} .
    \label{negative_condition} 
\end{equation}
In that case, ${\cal F}$ and $\varepsilon_p^{\rm std}$ achieve their maximally negative values for $\partial z_r/\partial z = 1/2$, 
\begin{equation}
      \left | \varepsilon_{p,min}^{\rm std} \right |  =  \kappa_T N_0^2 
      \left ( \frac{1}{4} - |\nabla_h z_r|^2 \right ) <  
      \frac{\kappa_T N_0^2}{4} ,
\end{equation}
which is bounded from above by $\kappa_T N_0^2/4$, which is laminar like in character and therefore extremely small relative to turbulent values. Nevertheless, because a negative $\varepsilon_p$ can potentially cause perturbations to grow with time, it seems of interest to estimate the time scale $\tau_{ape} = e_a/\varepsilon_p$ of the latter. Using the fact that the APE density scales as $N_0^2 \zeta^2/2$, where $\zeta = z-z_r$, this leads to
\begin{equation}
      \tau_{ape} = \frac{N_0^2\zeta^2/2}{\kappa_T N_0^2/4} = 
      \frac{2 \zeta^2}{\kappa_T} ,
      \label{time_scale_instability} 
\end{equation}
For illustration, using $\kappa_T = 10^{-7} {\rm m^2.s^{-1}}$ and $\zeta = 1 \,{\rm cm} = 10^{-2} {\rm m}$ yields $\tau = 2 \times 10^{-4}/10^{-7} = 2 \times 10^3 {\rm s} \approx 33 \,{\rm min}$, which is fast enough to be observable in the laboratory, at least in principle. Eq. (\ref{negative_condition}), which may be rewritten as $|\nabla_h \zeta| <1/2$, also requires that the horizontal scales of the unstable perturbations be larger than their vertical scales. A priori, perturbations with the required characteristics cannot occur spontaneously, they can only be created with some form of appropriate gentle mechanical stirring. These considerations appear to be consistent with the experimental observation of steps formation by \citet{Park1994}, who observed the smaller steps to grow faster than larger ones. However, these conclusions are only speculative at this stage and more research is needed to confirm or disprove the relevance of our results.

\subsection{Binary fluid: Spiceless case ${\cal M}=0$} 

In the double diffusive case but in the absence of spiciness, Eq. (\ref{epsilonp_final}) reduces to:
\begin{equation}
     \varepsilon_p = \textcolor{black}{K_{v0}} N_0^2 (z_r) 
     \left ( \frac{R_{\rho} - \tau }{R_{\rho} - 1 } \right ) 
     \left ( 1 - \frac{\partial_z z_r}{|\nabla z_r|^2} \right ) .
     \label{binary_no_spiciness} 
\end{equation}
This time, the sign of $\varepsilon_p$ is controlled by the sign of the product of ${\cal F}$ times the density-ratio dependent term within parentheses. Restricting attention to the case $\tau<1$ pertaining to thermosolutal convection, Table \ref{tab:Llt1} shows that $\varepsilon_p$ can be negative for all possible values of $R_{\rho}$, although not necessarily in both laminar and turbulent regimes. 

\begin{table}
  \begin{center}
\def~{\hphantom{0}}
  \begin{tabular}{c|c|c|c}
        & $-\infty < R_{\rho} \le \tau $ &  $\tau  < R_{\rho} \le 1$ & $1 < R_{\rho} < + \infty$ 
      \\[3pt] \hline
      ${\cal F} >0$   & $\epsilon_p > 0$ & \fbox{$\epsilon_p < 0$}  & $\epsilon_p > 0 $\\[3pt]
       ${\cal F} <0$   & 
       \fbox{$\epsilon_p < 0$}
       & $\epsilon_p > 0$ & \fbox{$\epsilon_p <0$} \\ \hline 
  \end{tabular}
  \caption{Sign of the APE dissipation rate $\varepsilon_p$ as a function of the density ratio $R_{\rho}$ and of the sign of ${\cal F} = |\nabla z_r|^2 - \partial_z z_r$ in the absence of spiciness (${\cal M}=0)$, in the case where $\tau  = \kappa_S/\kappa_T<1$ characteristic of salty water.}
  \label{tab:Llt1}
  \end{center}
\end{table}

\subsubsection{\textcolor{black}{Diffusive convection}: $\tau < R_{\rho} < 1$} 

In that case, $\varepsilon_p$ is negative only if ${\cal F}>0$, in which case the diapycnal component of the buoyancy flux is associated with up-gradient diffusion, as shown by \citet{Middleton2020}. Note that the range $\tau < R_{\rho} < 1$ is associated with \textcolor{black}{diffusive convection} instability only for inviscid flows. In reality, stability analysis shows that viscosity restricts the range of instabilities to the range 
\begin{equation}
         \frac{Pr + \tau}{Pr + 1} < R_{\rho} < 1 ,  
         \label{diffusive_convection_regime} 
\end{equation}
where $Pr = \nu/\kappa_T$ is the Prandtl number, with $\nu$ the molecular viscosity. For typical oceanic conditions, $\tau \approx 0.01$ and ${\rm Pr} \approx 7$, yields the restricted range $0.876 < R_{\rho}<1$.  

\subsubsection{`Salt' finger instability: $R_{\rho}>1$} 

According to classical linear stability analysis, salt fingers are expected to develop in the density ratio range
\begin{equation}
        1 < R_{\rho} < \frac{1}{\tau}, 
\end{equation}
e.g. \citet{Radko2013}. In that range, Table \ref{tab:Llt1} shows that APE dissipation can only be negative if ${\cal F}<0$, and hence that the appropriate expression for $\varepsilon_p$ is
\begin{equation}
       \varepsilon_p = \kappa_T N_0^2
       \left ( |\nabla z_r|^2 - 
       \frac{\partial z_r}{\partial z}\right ) 
       \left ( \frac{R_{\rho} - \tau}{R_{\rho}-1} \right ) 
       = \kappa_T N_0^2 {\cal F} 
       \left ( \frac{R_{\rho}-\tau}{R_{\rho}-1} \right ) .
       \label{epsilon_salt_fingers}
\end{equation}
Eq. (\ref{epsilon_salt_fingers}) shows that negative APE dissipation in that case corresponds to the amplification of the negative APE dissipation in a single component fluid multiplied by the following amplification factor 
\begin{equation}
        \frac{R_{\rho}-\tau}{R_{\rho}-1}  > 1  .
        \label{amplification_ratio} 
\end{equation}
Since the time scale for the growth of an unstable perturbation is a priori inversely proportional to $\varepsilon_p$, the expectation is that the same perturbations in a single component and double diffusive fluid will grow faster in the former than the latter. A key difference, however, is that in a double diffusive fluid, diffusive instabilities are able to grow spontaneously, whereas in a single component fluid their development requires initiation by means of external mechanical stirring of the right kind. 

Since ${\cal F}$ can only be negative for laminar-like conditions, it follows that density-compensated thermohaline variations need to develop in order for the APE dissipation to remain negative as the flow transitions to turbulence. As it happens, this is precisely what appears to happen in the salt-finger experiments of \citet{Middleton2020}, and as was previously advocated by \citet{Merryfield2000}. 

\begin{figure}
\center
\includegraphics[width=10cm]{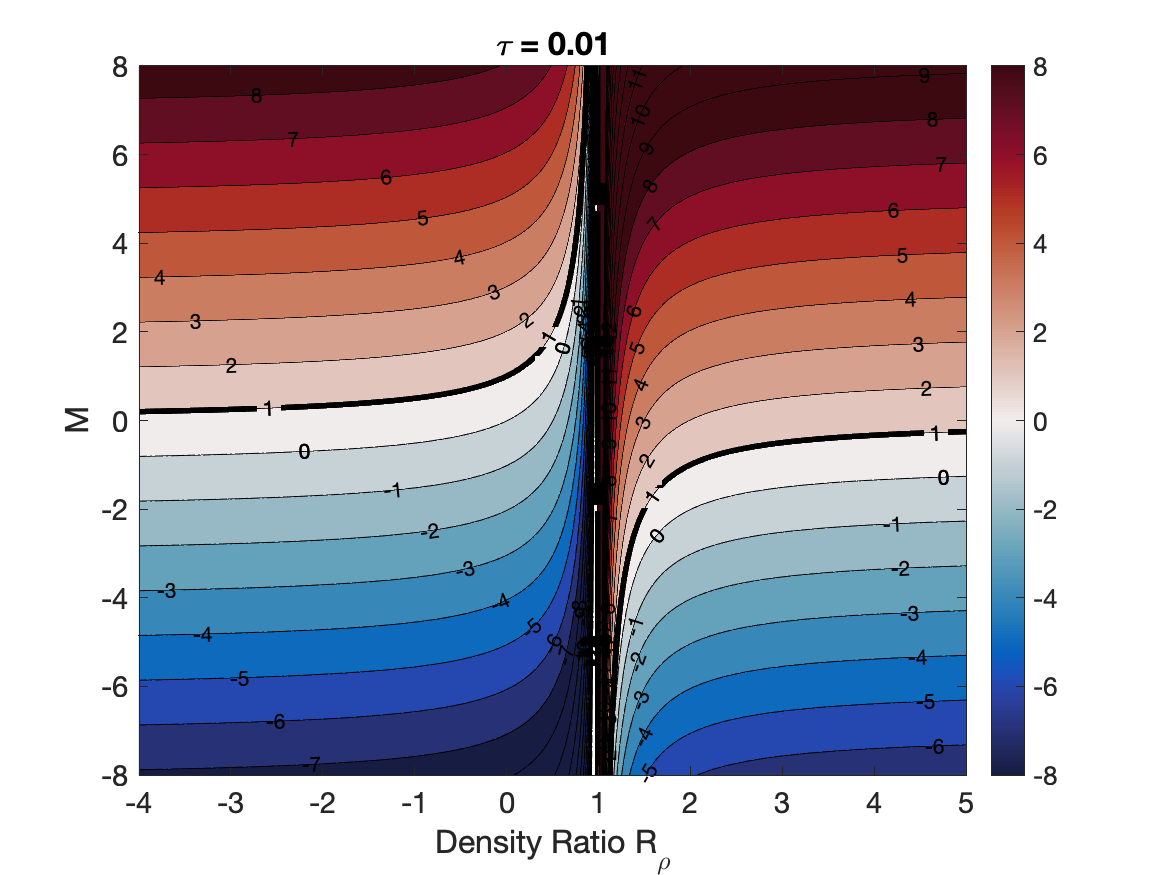}
\caption{The amplification factor $f(R_{\rho},\tau,{\cal M})$ (given by Eq. (\ref{amplification_factor})) as a function of the spiciness parameter ${\cal M}$ and density ratio $R_{\rho}$, for $\tau=0.01$}
\label{kt_to_ks_figures} 
\end{figure}

\subsection{Impact of spiciness anomalies on diffusive instability} 

Density-compensated thermohaline variations are a characteristic feature of ocean stratification in all regions and at all depths \citep{Tailleux2021}. Consequently, spiciness and a non-zero value of ${\cal M}$ are likely to be representative of the general situation. Assuming a prevalence of the turbulent regime, we can make the approximation $1 - \partial_z z_r/|\nabla z_r|^2 \approx 1$, and the relevant expression for $\varepsilon_p$ in the oceanic case appears as follows:

\begin{equation}
\varepsilon_p
= \textcolor{black}{K_{v0}} N_0^2(z_r) \left [ \frac{R_{\rho} - \tau }{R_{\rho}-1}
+ (1-\tau ) {\cal M} \right ] .
\label{APEdissipation_general}
\end{equation}

This equation shows that the single-component APE dissipation rate $\varepsilon_p^{std} = \textcolor{black}{K_{v0}}
N_0^2(z_r)$ is modulated by the factor:

\begin{equation}
f(R_{\rho},\tau,{\cal M}) = \frac{R_{\rho}-\tau}{R_{\rho}-1}
+ (1-\tau) {\cal M} .
\label{amplification_factor}
\end{equation}

To illustrate the effect of this factor, Figure \ref{kt_to_ks_figures} depicts $f(R_{\rho},\tau,{\cal M})$ as a function of $R_{\rho}$ and ${\cal M}$ for $\tau = \kappa_S/\kappa_T = 0.01$. A notable difference from the spiceless case is the possibility of negative $\varepsilon_p$ for a wide range of stratification, depending on the values of ${\cal M}$:

\begin{equation}
{\cal M} < - \frac{1}{1-\tau} \frac{R_{\rho}-\tau}{R_{\rho}-1} .
\end{equation}

This opens up various scenarios, ranging from complete inhibition of mixing ($f = 0$) to enhanced mixing ($f > 1$) and even diffusive instability ($f < 0$). In the turbulent case, the value of ${\cal M}$ approximates to:

\begin{equation}
{\cal M} = \frac{g}{2 \rho_{\star} N_0^2} \frac{(\nabla z_r\cdot \nabla \xi - \partial_z \xi)}{|\nabla z_r|^2 - \partial_z z_r} \approx
\frac{g}{2\rho_{\star} N_0^2} \frac{\nabla z_r\cdot \nabla \xi}{|\nabla z_r|^2}
= - \frac{\nabla \xi \cdot \nabla \rho}{2|\nabla \rho|^2} .
\end{equation}

This reveals that for ${\cal M}$ to be negative, spiciness and density gradients need to be positively correlated. Exploring the possibility and magnitude of such a situation requires a way to express in terms of the large-scale gradients of $\xi$ and $\rho$, which warrants further investigation beyond the scope of this study.

\section{Comparison of APE frameworks}
\label{interpretation} 

\citet{Holliday1981} and \citet{Andrews1981} had already shown the possibility of building Lorenz APE theory from a local principle by the time W95 published their study in the mid-1990s. However, the local APE framework was still rudimentary and had not been used for any concrete applications. This situation has improved in the recent years, as the local APE framework has progressed rapidly and achieved full maturity, being able to deal with a fully compressible multi-component stratified fluid \citep{Tailleux2018}. In this paper, we prefer the local APE framework to the global APE framework, because it is objectively simpler and more physical, even though this is not necessarily widely appreciated or acknowledged. The next subsection intends to reveal some of the unphysical aspects of the global APE framework that are commonly overlooked, but easily uncovered by the local APE framework. 

We first start by reviewing the main resuls of the global APE framework as first derived by W95. Following \citet{Lorenz1955}, W95 defined the APE of a fluid as the
potential energy of the actual state minus that of the adiabatically sorted state 
\begin{equation}
    E_a^{W95} = \int_{V} \rho g z \,{\rm d}V - \int_{V} \rho_r g z_r\,{\rm d}V
    = \int_{V}\rho g (z-z_r) \,{\rm d}V 
\end{equation}
and showed that its budget equation may be written in the form
\begin{equation}
     \frac{dE_a^{W95}}{dt} = \underbrace{\int_{V} \rho g w \,{\rm d}V}_{\Phi_z} 
     \,\,\underbrace{- g \oint_{S} 
     (z\rho -\psi) {\bf v}\cdot {\bf n} {\rm d}S}_{S_{APE}^{adv}} 
      + \underbrace{\kappa g \oint_{S} 
     (z - z_r) \nabla \rho \cdot {\bf n} \,{\rm d}S}_{S_{APE}^{diff}}  
     - (\Phi_d-\Phi_i) 
     \label{global_ape_budget} 
\end{equation}
where $\Phi_z$ represents the conversion with kinetic energy, $S_{ADV}^{adv}$ represents the 
advective flux of APE through the boundaries enclosing the control volume considered, $S_{APE}^{dif}$ represents the diffusive flux of APE through the same boundaries. The expressions for $\Phi_i$ and
$\Phi_d$ are given by
\begin{equation}
      \Phi_i = -\kappa g A (\overline{\rho}_{top} - \overline{\rho}_{bottom} )
\end{equation}
\begin{equation}
      \Phi_d = - \kappa g \int_{V} \frac{\partial z_r}{\partial \rho} |\nabla \rho |^2 \,{\rm d}V 
\end{equation}
where $A$ is the cross-section of the domain, assumed constant, the overbar denoting 
surface-average, while the quantity $\psi$ entering the expression of the advective flux is given by 
\begin{equation}
      \psi = \int^{\rho} z_r(\tilde{\rho},t) \,{\rm d}\tilde{\rho} .
\end{equation}
For the most part, the derivation of (\ref{global_ape_budget}) is straightforward, 
except for the demonstration that
\begin{equation}
     \int_{V} \rho \frac{\partial z_r}{\partial t} \,{\rm d}V = 0 .
\end{equation}
which is non-trivial. A key point of this paper concerns the proper interpretation of $\Phi_i$ and
$\Phi_d$. Following W95, $\Phi_d$ and $\Phi_i$ have been interpreted as representing two different
types of energy conversion, with BPE and IE respectively. The need for revisiting this interpretation is explained below.

The global APE budget (\ref{global_ape_budget}) is now compared with that obtained by 
integrating the local APE density budget. This is done here for a two-constituent fluid
with time-dependent Lorenz reference state, for which the local APE density takes the form
\begin{equation}
     e_a = e_a(S,\theta,z,t) =  -\int_{z_r}^z b(S,\theta,\tilde{z},t)\,{\rm d}\tilde{z} 
     = \frac{g}{\rho_{\star}} \int_{z_r}^z [ \rho(S,\theta) - \rho_0(\tilde{z},t) ] 
     \,{\rm d}\tilde{z} ,
     \label{ape_density_binary} 
\end{equation}
with the global APE being now defined as the volume-integrated APE density
\begin{equation}
     E_a = \int_{V} \rho_{\star} e_a \,{\rm d}V .
\end{equation}
\textcolor{black}{Because (\ref{ape_density_binary}) implies that 
$\rho_{\star} e_a = \rho g (z-z_r) + p_0(z,t) - p_0(z_r,t)$, it follows that
$E_a$ can be regarded as the sum of $E_a^{W95}$ plus an additional term 
related to the work against the vertical reference pressure gradient as follows:
\begin{equation}
    E_a = E_a^{W95} + \int_{V} [ p_0(z,t) - p_0(z_r,t) ] \,{\rm d}V .
    \label{link_between_ape_definitions} 
\end{equation}
Eq. (\ref{link_between_ape_definitions}) shows that the validity of the global
definition of APE $E_a^{W95}$ 
requires the second term pertaining to the work against the reference pressure
gradient to vanish. This is easily verified to be the case if the volume $V$ 
enclosing the fluid remains unchanged under the isochoric adiabatic 
re-arrangement ${\bf x} \rightarrow {\bf x}_r$ used to construct $\rho_0(z,t)$,
as then $\partial (x,y,z)/\partial (x_r,y_r,z_r) = 1$, 
${\rm d}V = {\rm d} V_r$, $V_r = V$, thus ensuring that
\begin{equation}
     \int_{V} p_0(z_r,t) \,{\rm d}V = \int_{V_r} p_0(z_r) \,{\rm d}V_r  
     = \int_{V} p_0(z,t) \, {\rm d}V .
\end{equation}
as requested, e.g., \citet{Roullet2009}, \citet{Tailleux2013b} and \citet{Winters2013} for instance. 
However, the identity $E_a = E_a^{W95}$ is not expected to necessarily hold for
more complex situations, although the precise circumstances under which this 
happens have yet to be fully understood and discussed in the literature. 
} 

For a time-dependent Lorenz reference state, the local APE budget (\ref{ape_density_evolution})
derived earlier needs is easily shown to be given by
\begin{equation}
    \frac{\partial e_a}{\partial t} = 
    - {\bf v}\cdot \nabla e_a - b w 
    - \nabla \cdot {\bf J}_a - \varepsilon_p 
    - \frac{g}{\rho_{\star}} \int_{z_r}^z 
    \frac{\partial \rho_0}{\partial t} (\tilde{z},t) \,{\rm d}\tilde{z}  
    \label{local_ape_budget_with_varying_rho0}
\end{equation}
where ${\bf J}_a = -(z-z_r){\bf J}_b$ and $\varepsilon_p = {\bf J}_b \cdot \nabla (z-z_r)$ as before.
By taking the volume integral of (\ref{local_ape_budget_with_varying_rho0}), the evolution 
equation for the global APE $E_a$ is easily shown to be
\begin{equation}
     \frac{dE_a}{dt}  
      = 
            \underbrace{-\int_{V} \rho_{\star} b w \,{\rm d}V}_{-\Phi_z}       
      \underbrace{- \oint_{S} \rho_{\star}  e_a {\bf v} \cdot {\bf n} dS}_{S_{APE}^{adv}} 
      + \underbrace{\oint_{S} \rho_{\star} (z-z_r) {\bf J}_b \cdot {\bf n}\,{\rm d}S}_{S_{APE}^{diff}}
      -\underbrace{\int_{V} \rho_{\star} \varepsilon_p \,{\rm d}V}_{\Phi_d-\Phi_i}  ,
      \label{volume_integrated_ape_boussinesq}
\end{equation}
where the expressions for $\Phi_i$ and $\Phi_d$ are 
\begin{equation}
\begin{split} 
     \Phi_i = &  - \int_{V} \rho_{\star} \varepsilon_{p,lam} 
     = \rho_{\star} g A \left [ \alpha \kappa_T \Delta \theta 
     -  \beta \kappa_S \Delta S  \right ] \\
     = & \rho_{\star} g A \kappa_T \beta \Delta S 
     \left [ \frac{\alpha \Delta \theta}{\beta \Delta S} - \frac{\kappa_S}{\kappa_T} \right ] 
     = \rho_{\star} g A \kappa_T \beta \Delta S \left ( R_{\rho}^{bulk} - \tau \right ) 
\end{split} 
\label{phi_i_expression} 
\end{equation}
\begin{equation}
     \Phi_d = \int_{V} \rho_{\star} \varepsilon_{p,tur} =
     \int_{V} g (\alpha \kappa_T \nabla \theta - \beta \kappa_S \nabla S ) 
     \cdot \nabla z_r\rho_{\star} \,{\rm d}V  
\end{equation}
where $\Delta \theta = \overline{\theta}_{top} - \overline{\theta}_{bottom}$ and $\Delta S = \overline{S}
_{top} - \overline{S}_{bottom}$,
while $R_{\rho}^{bulk}$ is a bulk density ratio. For canonical salt finger instability, \textcolor{black}{$1< R_{\rho}^{bulk} < 1/\tau$, in which case (\ref{phi_i_expression}) predicts} 
$\Phi_i>0$ as required to act as a source of APE. Note that the last term in (\ref{local_ape_budget_with_varying_rho0}) may be written in the form $F(z,t)-F(z_r,t)$, with $F$ such that $\partial F/\partial z(z,t) = \partial \rho_0/\partial t(z,t)$. \textcolor{black}{As a result, its volume integral must vanish 
whenever the second term of (\ref{link_between_ape_definitions}) also vanishes
for the same reason, as explained above.}

 \textcolor{black}{\citet{Winters2013} assumed that when $E_a$ equals $E_a^{W95}$,
their respective budgets should be equivalent and described by the equations for
an open domain derived previously 
by \citet{Winters1995}. This view is clearly incorrect, 
however, because the budgets of $E_a$ and $E_a^{W95}$ are closely related to the local budgets of their integrands, but these are not equivalent as established above. Physically, the non-equivalence of the budgets for $E_a$ and $E_a^{WP5}$ is due to the integrand of $E_a^{W95}$ lacking a vital piece of information about the energetics of stratified fluids compared to $E_a$, namely the fact that when a fluid parcel moves from its reference position $z_r$ to its actual position, it not only has to work against gravity, but also against the vertical reference pressure gradient. More generally, \citet{Tailleux2013b} shows that work against buoyancy forces actually involves three terms: 1) one associated with changes in gravitational potential energy; 2) one associated with changes in internal energy; 3) one associated with work against the reference pressure gradient. Even if the latter term integrates to zero over the whole fluid domain, it is still necessary to retain it for correctly predicting the advective flux of APE through the boundaries of open domains.
Two key differences characterise the non-equivalence of the budgets for $E_a$ and
$E_a^{W95}$:} 
1) in W95, the reversible conversion with kinetic energy $\Phi_z$ is associated with the density flux $\rho g w$ but with the buoyancy flux $-b w$ in the local APE framework; 2) in W95, the advective flux of APE through boundaries $S_{APE}^{adv}$ is associated with the flux of the bizarre quantity
$g (z\rho-\psi)$ but with the flux of APE density $e_a$ in the local framework, 
which makes much more sense. The unphysical character of the advective flux of APE in W95 highlights
the well known problematic character of the global APE framework for regional studies, which was one of the key motivations for the quest of a local available energy framework \citep{Tailleux2013}. 

\textcolor{black}{The unphysical character of the advective flux of APE in \citet{Winters1995} global APE budget for open domains is a problematic and
can only be corrected by using the local APE framework.
The issue is vital, because recent studies of local energetics have
demonstrated the key importance of the lateral advection of APE density for 
understanding important physical processes, such as storm track dynamics \citep{Novak2018,Federer2024}, 
tropical cyclone intensification \citep{Harris2022}, or the energetics of
the thermohaline circulation \citep{Gregory2011,Sijp2012}. Importantly, it is 
worth noting that $E_a$ and $E_a^{W95}$ are also non-equivalent numerically. 
Indeed, in the global
framework, the APE of the fluid is estimated as a small residual between two very
large terms, which is ill-conditioned and prone to large numerical errors. 
In the local framework, however, the APE of a fluid is estimated as the sum of
small non-negative numbers, which is numerically robust and very accurate. 
As a result, the field would greatly benefit from switching to the local APE
framework, as the continued use of the global APE framework only serves to 
perpetuate generally unsound and inferior practices of limited usefulness.} 

\section{Reference state variables} 
\label{reference_state_variables}

The practical implementation of the above theoretical results require the computation of Lorenz reference density profile $\rho_0(z,t)$, as well as of the associated isopycnal temperature and salinity averages necessary for defining spiciness. Here, we provide the mathematical foundations for defining the latter in terms of probability density functions (p.d.f.), which allows one to compute these in practice without the use of sorting. These results pertain to arbitrary domains and unify the p.d.f approaches of \citet{Tseng2001} and \citet{Saenz2015}, while also making use of previous ideas by \citet{Hochet2019b} and \citet{Hochet2021}. 

\smallskip 

\subsection{Reference position $z_r(\rho,t)$ and density profile $\rho_0(z,t)$} 

The classical interpretation of Lorenz reference state as an isochoric re-arrangement of fluid parcels is easy to understand and implement algorithmically. However, it does not easily lend 
itself to mathematical analysis. Here, we derive expressions for defining and constructing Lorenz reference density profile $\rho_0(z,t)$ and its inverse function $z_r(\rho,t)$ [such that $\rho_0(z_r(\rho,t),t) = \rho$ at all times]
in a way that greatly facilitates the study of its mathematical properties. Our approach is not
entirely new, as elements of it can be found in \citet{Winters1995} or in the p.d.f. approaches
of \citet{Tseng2001}, \citet{Saenz2015} or \citet{Hochet2019b,Hochet2021}, but is useful to 
introduce the approaches discussed next. To that end, let us
assume that the domain containing the fluid is bounded at the top by a rigid lid at $z=0$ and at its bottom by a spatially varying bottom topography at $z=-H(x,y)$. We denote by $\hat{V}(z)$ the volume between an arbitrary depth $z$ and the surface, and by $A(z) = -{\rm d}V/{\rm d}z(z)$ 
the cross-sectional area of the domain at the depth $z$.

\begin{figure}
\center
\includegraphics[width=10cm]{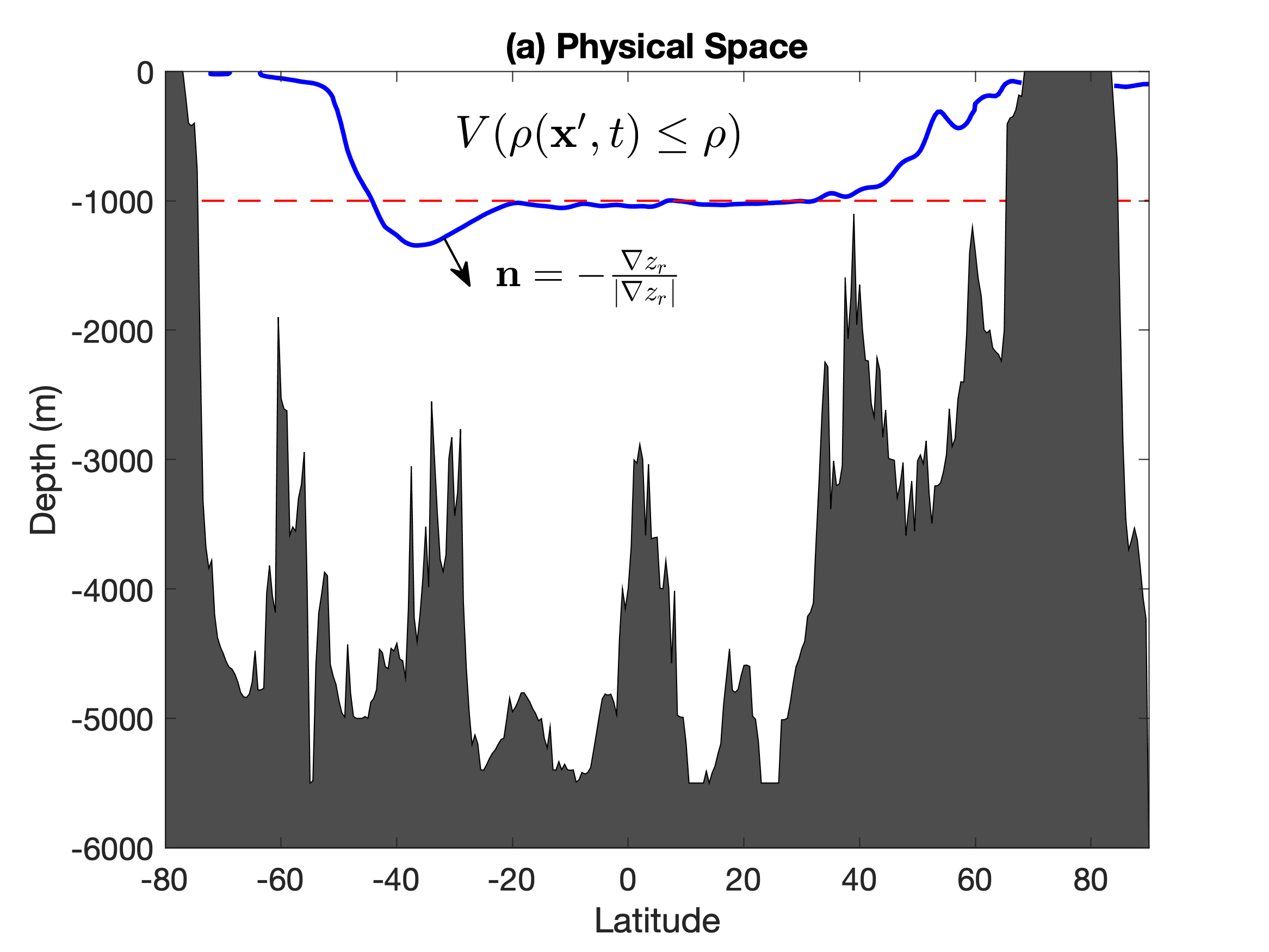}
\includegraphics[width=10cm]{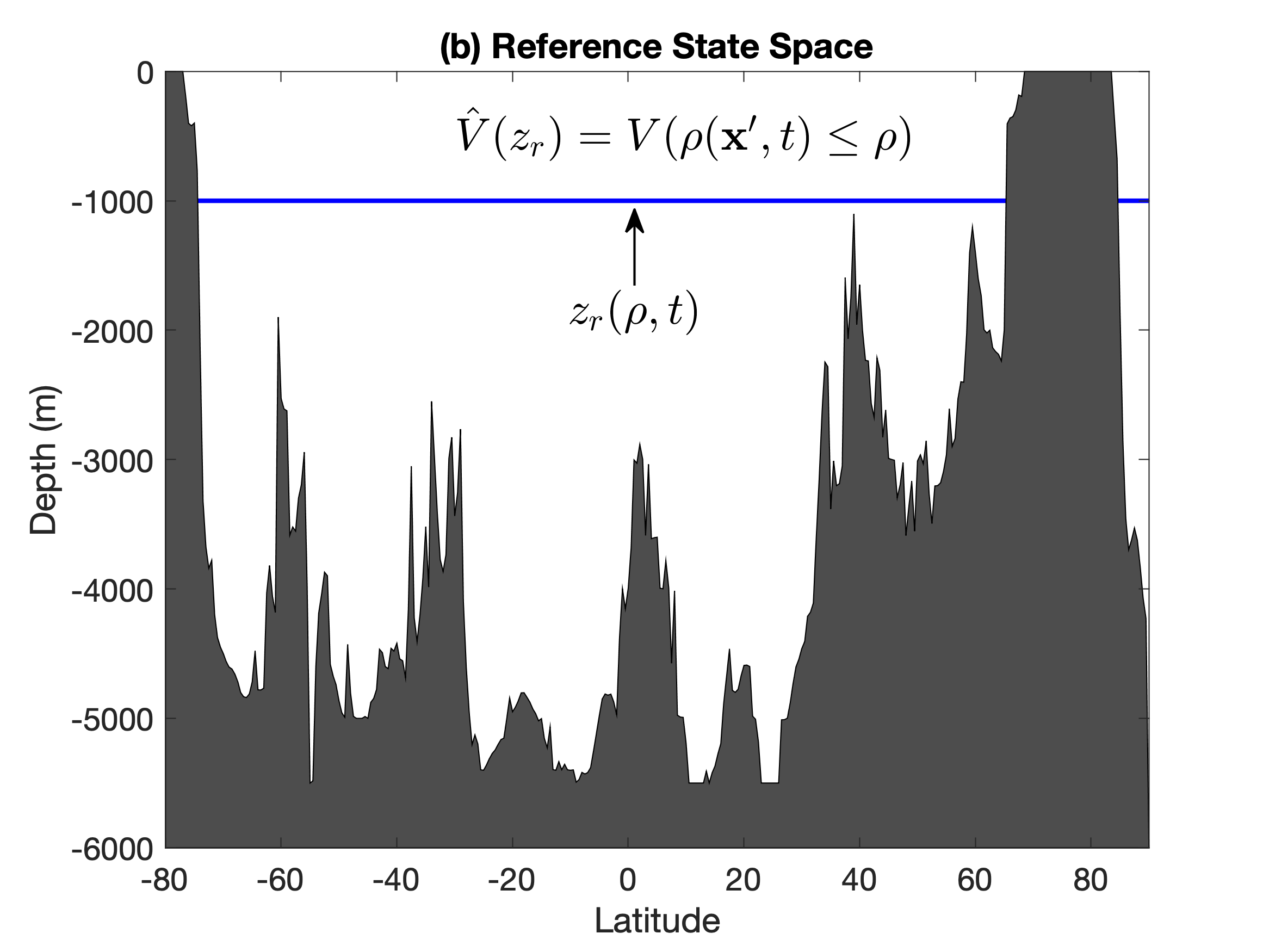}
\caption{Mapping of physical space (a) onto reference state space (b).}
\label{physical_vs_reference_space}
\end{figure}

Physically, the reference depth $z = z_r(\rho,t)$ of a fluid parcel must be such that the volume of water between $z_r(\rho,t)$ and the surfafe, viz., 
\begin{equation}
    \hat{V}(z_r) = \int_{z_r}^0 A(z)\,{\rm d}z 
\end{equation}
must equal the volume of water in physical space made up of all the fluid parcels of densities
less than $\rho$, viz., 
\begin{equation}
     V(\rho,t) = \int_{V(\rho({\bf x}',t) \le \rho)}  {\rm d}V' 
     = \int_{V}  H(\rho - \rho({\bf x}',t)) \,{\rm d}V' 
\end{equation}
where $H$ denotes the Heaviside step function defined as in \citet{Winters1995},
\[
   H(\rho({\bf x},t) - \rho({\bf x}_0,t)) = \left\{
    \begin{array}{ll}
       0, &\rho ({\bf x},t) < \rho({\bf x}_0,t)\\[2pt]
      \frac{1}{2},      & \rho({\bf x},t) = \rho({\bf x}_0,t) \\ [2pt]
      1, & \rho({\bf x},t) > \rho({\bf x}_0,t)      
      \end{array} \right. . 
\]
Equating the two volumes $\hat{V}(z_r) = V(\rho,t)$ with $V(\rho,t)$ yields an implicit 
equation for $z_r(\rho,t)$, which may be inverted to yield 
\begin{equation}
      z_r(\rho,t) = \hat{V}^{-1}[V(\rho,t)] ,
      \label{zrho_definition}
\end{equation}
which in the case where $A(z) = {\rm constant}$ reduces to 
\begin{equation} 
    z_r(\rho,t) = -\frac{V(\rho,t)}{A} 
\end{equation}
as previously derived by \citet{Winters1995}. As an illustration, Fig. \ref{physical_vs_reference_space} (a) shows the constant Lorenz reference depth $z_r(\rho,t) = -1000\,{\rm m}$ for a realistic ocean, with panel (b) showing the corresponding position in
reference state space. Construction of Lorenz reference state for a realistic ocean has been discussed in \citet{Saenz2015,Tailleux2016b,Tailleux2021}. Once $z_r(\rho,t)$ has been determined for a number of target densities, the reference density profile $\rho_0(z,t)$ may be obtained simply
by solving the implicit relation $\rho_0(z_r(\rho,t),t) = \rho$ at a number of discrete reference
depths, from which an interpolator may be constructed to infer the reference density for other values. Alternatively, using the fact that $\rho = \rho_0(z_r,t)$ by definition, one may also solve the equation $\hat{V}(z_r) = V(\rho,t)$ in the form $\hat{V}(z_r) = V(\rho_0(z_r,t),t)$ to infer $\rho_0(z_r,t)$ at a number of predetermined target reference depths. 

Note that by differentiating the equality $\hat{V}(z_r) = V(\rho,t) = V(\rho_0(z_r,t),t)$ with respect to $z_r$, the following useful expression for 
$\partial \rho_0/\partial z_r$ may be obtained,
\begin{equation}
     \frac{\partial \rho_0}{\partial z}(z_r,t) 
     = -\left (\frac{\partial V}{\partial \rho} \right ) ^{-1} 
     A(z_r) .
     \label{useful_relation} 
\end{equation}
The above expressions establish that the practical computation of $z_r(\rho,t)$ and $\rho_0(z,t)$ 
can be achieved without resorting to a sorting algorithm.
Indeed, all one has to do is to define a number of target densities $\rho_i, {i=1,\cdots,N}$ at which to compute the corresponding volumes $V(\rho_i,t)$. Having defined the inverse function $\hat{V}^{-1}(z)$, one may simply compute the corresponding target reference depths $z_r(\rho_i,t)$ at which $\rho_0(z,t)$ may be evaluated. Once all quantities have been determined at the discrete target depths, interpolators can be constructed to find values at other points. 

\subsection{Construction of isopycnal mean profiles} 

\begin{figure}
    \center
    \includegraphics[width=10cm]{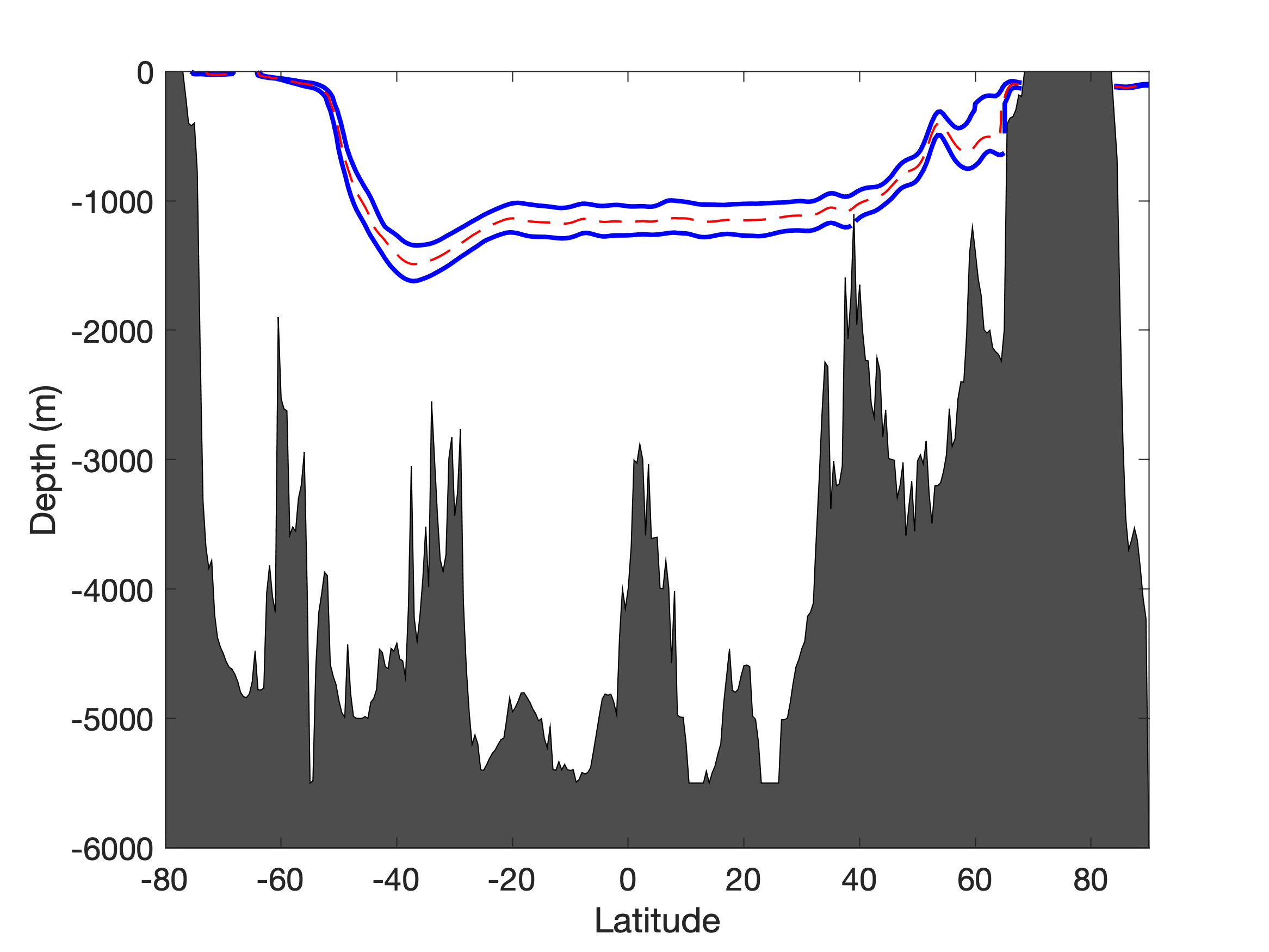}
    \caption{Schematics of an isopycnal surface (dashed-red line) bounded by two nearby isopycnal surfaces (thick black lines) defining the volume of fluid used to define the thickness-weighted average value of any arbitrary tracer for the isopycanl surface considered.}
    \label{twa_graphics}
\end{figure}

Physically, the isopycnally-averaged temperature and salinity profiles $\theta_0(z,t)$ and
$S_0(z,t)$ needed to isolate the 
passive spice component need to be obtained as thickness-weighted isopycnal averages, 
as illustrated in Fig. \ref{twa_graphics}, as this is the only physically meaningful way to 
define an isopycnal average \citep{Young2012}. Mathematically, $\theta_0(z,t)$ can be 
defined as a solution of the following equation: 
\begin{equation}
     \int_{z_r(\rho,t)}^0 A(z)\theta_0(z,t) \,{\rm d}z = 
     \underbrace{\int_{V(\rho,t)} \theta({\bf x},t) \,{\rm d}V}_{H(\rho,t)}  .
     \label{eq_defining_theta0}
\end{equation}
Physically, the right-hand side of (\ref{eq_defining_theta0}) represents the volume integrated
temperature over the body of water made up of fluid parcels with densities less than $\rho$,
which we denote $H(\rho,t)$. The right-hand side is the volume integral of the thickness-weighted 
isopycnally averaged temperature between $z_r(\rho,t)$ and the surface. An explicit expression
for $\theta_0(z,t)$ at $z=z_r$ may be obtained by differentiating (\ref{eq_defining_theta0}) 
with respect to $z_r$, which leads to
\begin{equation}
     - A(z_r) \theta_0(z_r,t) = \frac{\partial H}{\partial \rho} 
     \frac{\partial \rho_0}{\partial z_r}(z_r,t) \qquad
     \Rightarrow \qquad \theta_0(z_r,t) = \left ( \frac{\partial V}{\partial \rho}  
     \right )^{-1} \frac{\partial H}{\partial \rho} 
     \label{theta0_definition} 
\end{equation}
which made use of (\ref{useful_relation}). 

\begin{figure}
\center
\includegraphics[width=6.5cm]{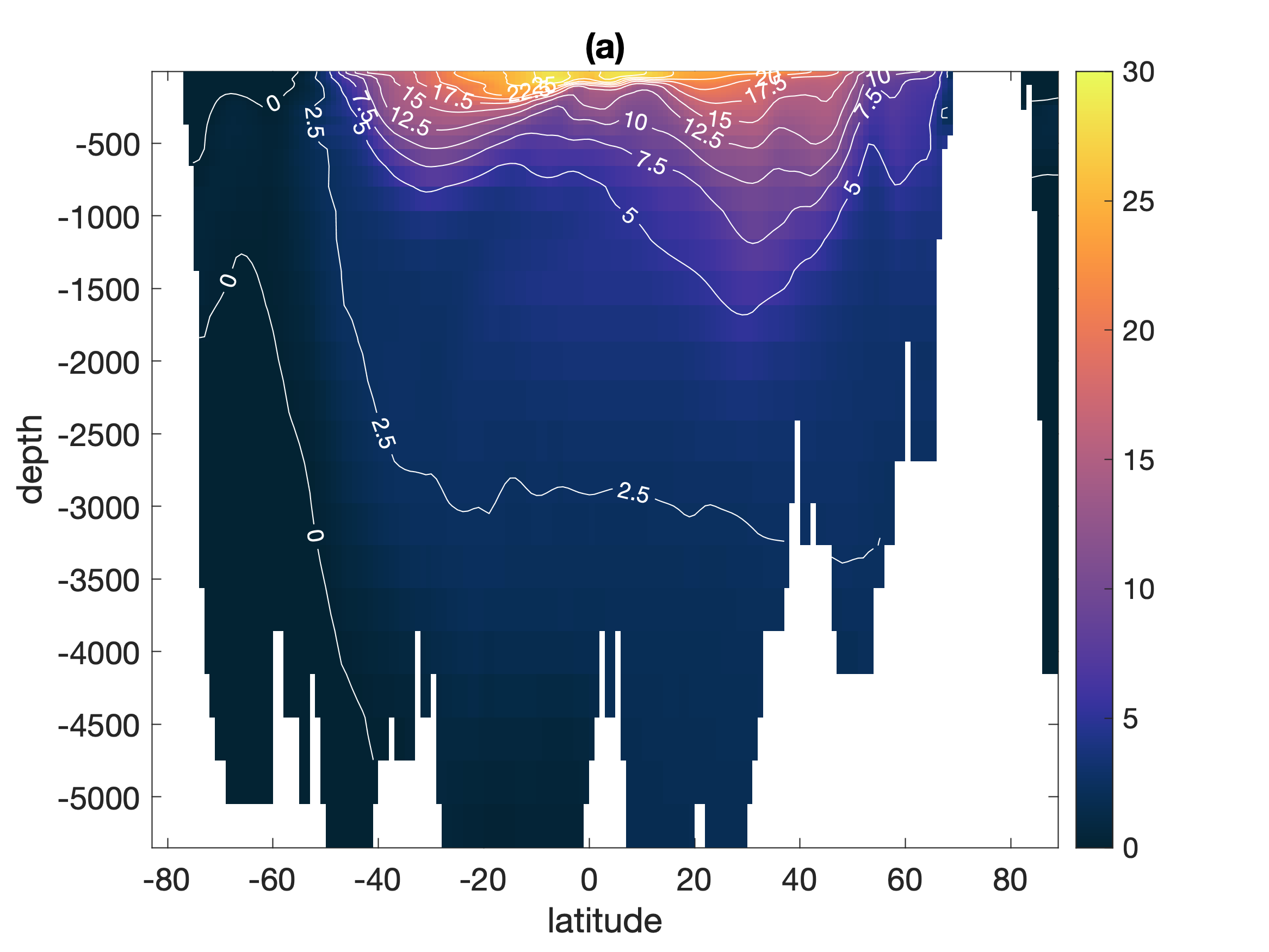}
\includegraphics[width=6.5cm]{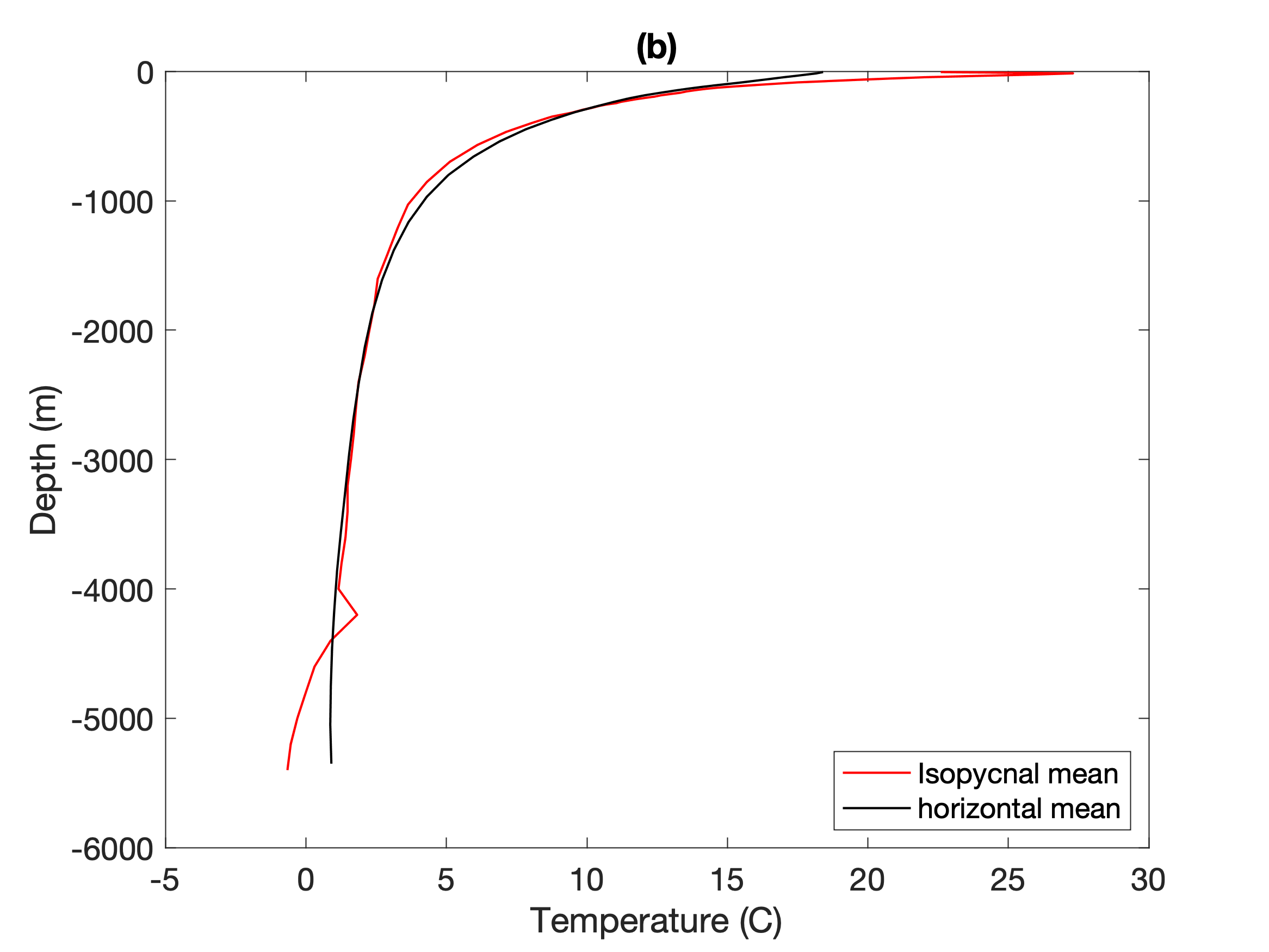}
\includegraphics[width=6.5cm]{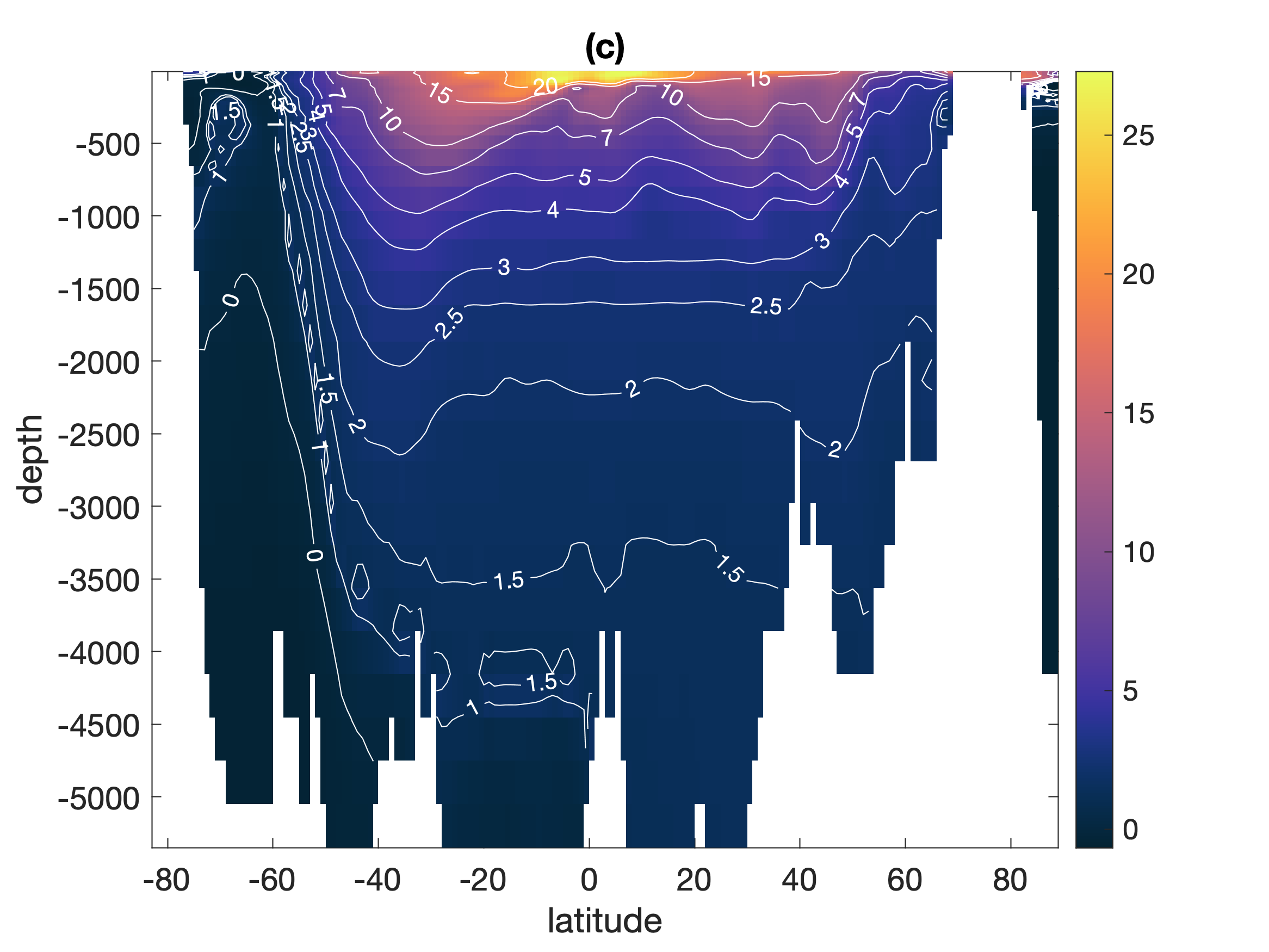}
\includegraphics[width=6.5cm]{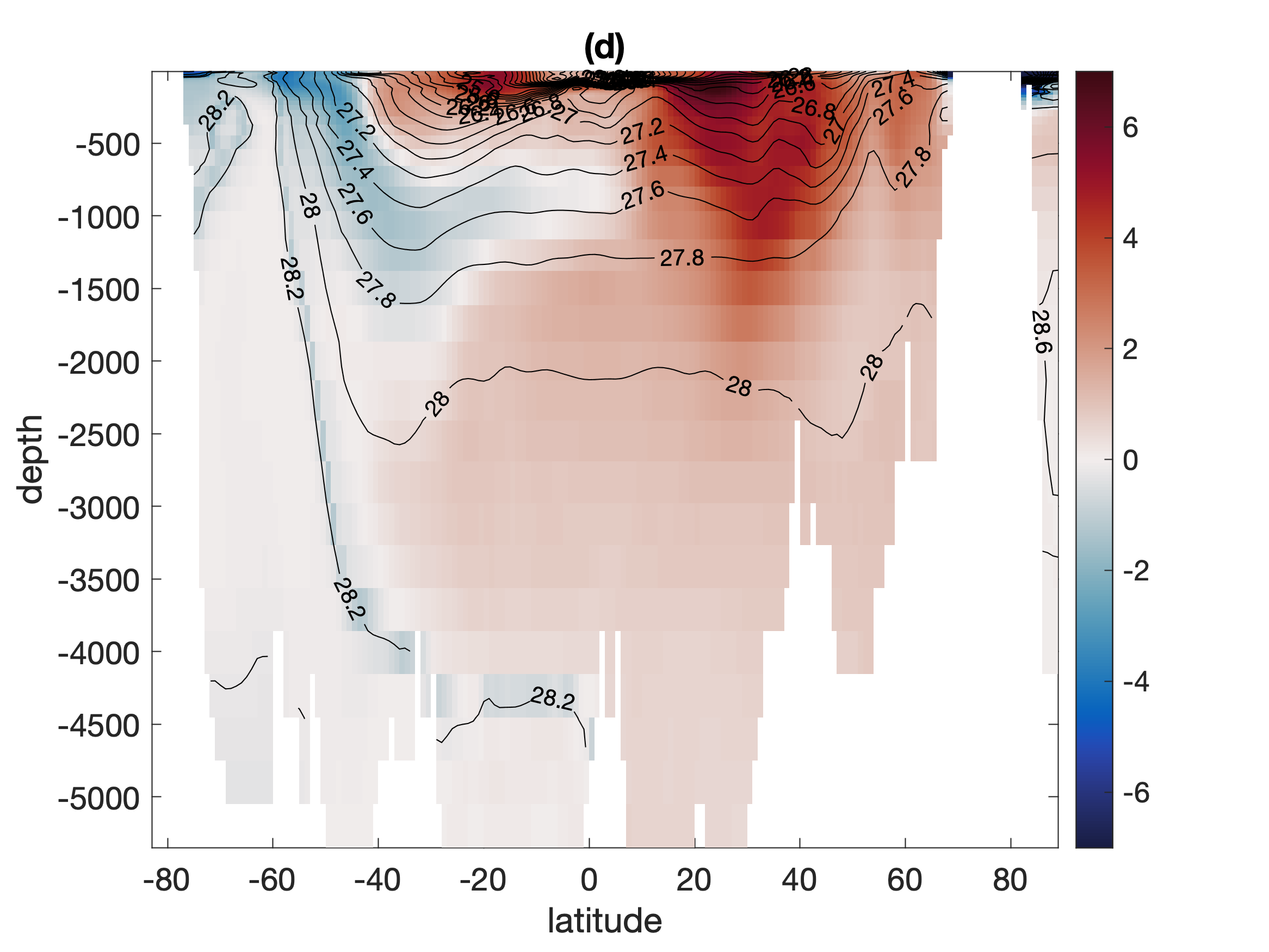} 
\caption{(a) Latitude/depth section of Conservative Temperature along $30^{\circ} W$ in the 
Atlantic ocean; (b) Global isopycnal mean (red line) profile used for the spice/heave 
decomposition of Conservative Temperature shown in (c) and (d). For comparison,
horizontal mean profile (black line) is also depicted. 
(c) Active heave component of Conservative Temperature along same
$30^{\circ} W$ section as in (a). (d) Passive spice component of Conservative Temperature along
same $30^{\circ} W$ section as in (a). Black contour lines represent constant Lorenz reference
density surfaces labelled in terms of thermodynamic neutral density $\gamma^T$.}
\label{spice_heave_decomposition} 
\end{figure}

Eq. (\ref{theta0_definition}) is easily implemented in practice,
\textcolor{black}{at least for the type of data considered here.}
As an illustration, 
Fig. \ref{spice_heave_decomposition} shows a decomposition of temperature into active and
passive components. Specifically, Fig. \ref{spice_heave_decomposition} (a) shows a meridional
section of Conservative Temperature along $30^{\circ} W$ in the Atlantic Ocean. 
Panel (b) shows the thickness-weighted isopycnal temperature average profile compared with the
more standard horizontal mean profile. The thickness-weighted averaged temperature is illustrated
in Panel (c), while Panel (d) illustrates the passive spice component, obtained as
$\theta_{spice} = \theta - \theta_0(z_r,t)$. Our approach to defining spice, based on 
\citet{Tailleux2021}, is strongly supported by the fact that the patterns in panel (d) all
appear to coincide with known ocean water massses in the Atlantic Ocean, namely 
Antarctic Bottom Water (AABW), North Atlantic Deep Water (NADW), Antarctic Intermediate Water
(AAIW), and Mediterranean Intermediate Water (MIW).

\subsection{Evolution of the reference density profile} 

By definition, $\rho_0(z,t)$ must coincide with its thickness-weighted isopycnal average and therefore satisfy (\ref{eq_defining_theta0}), viz., 
\begin{equation}
     \int_{z_r}^0 A(z) \rho_0(z,t) \,{\rm d}z 
     = \underbrace{\int_{V(\rho,t)} \rho({\bf x}',t) \,{\rm d}V'}_{R(\rho,t)} .
     \label{twa_density} 
\end{equation}
We now show that this equation can be manipulated to yield an exact expression for the evolution equation satisfied by $\rho_0(z,t)$ as as well as for the effective diffusivity first introduced by \citet{Winters1996} and \citet{Nakamura1996}. For simplicity, we focus only on the case where the buoyancy flux vanishes through all solid boundaries as well as the surface, but the derivation can easily be extended to account for a non-zero surface buoyancy flux. 

To obtain the sought-for evolution equation for $\rho_0$, let us take the time derivative of (\ref{twa_density}), which yields
\begin{equation}
    \int_{z_r}^0 A(z) \frac{\partial \rho_0}{\partial t} 
    (z,t) \,{\rm d}z = \int_{V(\rho,t)} 
    \frac{\partial \rho}{\partial t}({\bf x}',t)\,{\rm d}V' .
    \label{intermediate_result} 
\end{equation}
Now, using the fact that the evolution equation for density may be written as
\begin{equation}
     \frac{\partial \rho}{\partial t} = - {\bf v}\cdot \nabla \rho + \nabla \cdot 
     (\rho_{\star} {\bf J}_b/g) , 
\end{equation}
(\ref{intermediate_result}) may be rewritten as follows
\begin{equation}
\begin{split} 
    \int_{z_r}^0 A(z) \frac{\partial \rho_0}{\partial t} (z,t)\,{\rm d}z 
   = &  \int_{V(\rho,t)}  
    [ \nabla \cdot ( \rho_{\star} {\bf J}_b/g)  - {\bf v}\cdot \nabla \rho ]\,{\rm d}V' \\
    = & \oint_{S_{\rho}} (\rho_{\star}/g) {\bf J}_b \cdot {\bf n} \, {\rm d}S 
    = \oint_{S_{\rho}} (\rho_{\star}/g) {\bf J}_b \cdot \frac{\nabla \rho}{|\nabla \rho|} 
    \,{\rm d}S \\
    = &  \frac{\partial \rho_0}{\partial z}(z_r,t) \oint_{S_{\rho}} 
    (\rho_{\star}/g) \frac{{\bf J}_b \cdot \nabla z_r}{|\nabla \rho|} \,{\rm d}S \\
    = & - \frac{\partial \rho_0}{\partial z}(z_r,t) \oint_{S_{\rho}} 
    (\rho_{\star}/g) \frac{\varepsilon_{p,tur}}{|\nabla \rho|} \,{\rm d}S .
\end{split} 
\label{evolution_equation_first_step} 
\end{equation}
In the second line, $S_{\rho}$ refers to the isopycnal surface $\rho={\rm constant}$, meaning that the outward normal unit vector ${\bf n} = \nabla \rho/|\nabla \rho| = -\nabla z_r/|\nabla z_r|$ as schematically illustrated in Fig. \ref{physical_vs_reference_space} (a). To go from the second to third line we used the fact that as $\rho = \rho_0(z_r,t)$, we have $\nabla \rho = \partial \rho_0/\partial z_r(z_r,t) \nabla z_r$. To go from the third to fourth line, we used the fact that by definition ${\bf J}_b\cdot \nabla z_r= -\varepsilon_{p,tur}$. 

By defining the reference squared buoyancy profile as
\begin{equation}
     N_0^2 (z_r,t) = -\frac{g}{\rho_{\star}} 
     \frac{\partial \rho_0}{\partial z} (z_r,t)
\end{equation}
and the locally-defined effective diffusivity as 
\begin{equation}
    \kappa_{\rm eff}  = \frac{\varepsilon_{p,turb}}{N_0^2 (z_r)} ,
\end{equation}
(\ref{evolution_equation_first_step}) may be rewritten as 
\begin{equation}
    \int_{z_r}^0 A(z) \frac{\partial \rho_0}{\partial t} (z,t) \,{\rm d}z 
    =  - A(z_r) K_{\rm eff} \frac{\partial \rho_0}{\partial z_r} (z_r,t)  = \frac{\partial R}{\partial t} ,
    \label{evolution_equation_for_rho0_second_step}
\end{equation}
where the globally defined effective diffusivity is related to the locally-defined one by
\begin{equation}
    K_{\rm eff} = \frac{1}{A(z_r)} \oint_{S_{\rho}} 
    \frac{\kappa_{\rm eff}}{|\nabla z_r|} \,{\rm d}S  .
    \label{effective_diffusivity_definition}
\end{equation}
as previously established by \citet{Hochet2019}.
The final step of the derivation consists in differentiating 
(\ref{evolution_equation_for_rho0_second_step}) with respect to $z_r$, which yields
\begin{equation}
     \frac{\partial \rho_0}{\partial t} (z_r,t) 
     = \frac{1}{A(z_r)} \frac{\partial }{\partial z_r} 
     \left [ A(z_r) K_{\rm eff} \frac{\partial \rho_0}{\partial z_r} \right ] (z_r,t) .
     \label{diffusion_equation} 
\end{equation}
Note that in practice, $K_{\rm eff}$ can also 
be determined by inverting the diffusion equation (\ref{diffusion_equation}) from the knowledge of the temporal variations of the sorted density profile $\rho_0(z,t)$. Alternatively, one may combine Eqs. (\ref{useful_relation}) and (\ref{evolution_equation_for_rho0_second_step}) to verify that the globally-defined effective diffusivity may be written as
\begin{equation}
     K_{\rm eff}(z_r,t) = \frac{1}{A^2(z_r)} 
     \frac{\partial V}{\partial \rho} (\rho,t)
     \frac{\partial R}{\partial t} (\rho,t) 
\end{equation}
which in principle allows one to calculate $K_{\rm eff}$ using simple binning techniques. The above results establish that the effective diffusivity only depends on the turbulent part $\varepsilon_{p,tur}$ of $\varepsilon_p$, consistent with 
established results.

\section{Summary, discussion, and conclusions}
\label{discussion_section} 
Double diffusive instabilities are mysterious as they can seemingly develop from a state with zero initial mechanical energy, without the need for any external mechanical source(s) of energy (except for that of the initial perturbation triggering the instability). In the context of APE theory, such instabilities are possible only if the mechanical energy KE+APE can grow at the expense of BPE. 
\textcolor{black}{Since viscous dissipation $\varepsilon_k$ is always positive,
this can only happen when the APE dissipation rate $\varepsilon_p$ is negative.} Therefore, it is more logical and physically plausible to consider the sign of the APE dissipation rate $\varepsilon_p = {\bf J}_b \cdot \nabla (z-z_r)$, rather than the sign of the diapycnal component of the buoyancy flux proposed by \citet{Middleton2020}, as the most fundamental criterion for characterising double diffusive instabilities. In the context of \citet{Winters1995} framework, this is equivalent to associating the development of double diffusive instabilities with the sign of $\Phi_d-\Phi_i$ rather than that of $\Phi_d$ only. This is justified by the fact that $\Phi_d$ is only negative for diffusive convection instability 
but not for salt finger instability, while $\Phi_d-\Phi_i$ is negative for both. 

\textcolor{black}{While our current findings unequivocally acknowledge the existence 
and necessity of negative APE dissipation in single- and two-component stratified fluids, the precise underlying mechanisms remain elusive. 
The controversy primarily revolves
on the nature of background potential energy (BPE) 
and of the energy conversion type between APE and BPE. 
The primary challenge stems from the fact that, in fully compressible fluids, APE dissipation naturally 
appears as a conversion between APE and the internal energy component of BPE, akin to viscous dissipation \citep{Tailleux2009,Tailleux2013c}. However, this
interpretation is not the most intuitive within standard Bousinesq models, as 
the BPE lacks an explicit link to internal energy, thereby complicating our 
understanding of its energetics. Nevertheless, the crucial role of internal energy
in understanding the energetics of turbulent stratified fluids is underscored by the term $g z D\rho/Dt$, which approximates the 
compressible work $p D\upsilon/Dt$ in Boussinesq models \citep{Young2010,Tailleux2012}, implying 
a conversion with internal energy. While some may find considering thermodynamics
as central to the understanding of the energetics of turbulent stratified fluids unpalatable, it is notheworthy that, in the current state of knowledge, there does
not appear to be any other way to explain the occurrence of negative APE dissipation in single-component fluids or doubly stable two-component fluids. 
Indeed, existing explanations for the release of BPE into APE by 
\citet{Smyth2005} and \citet{Ma2021} have linked it to the `APE' associated
with a destabilising stratifying tracer, but in addition to lacking rigorous
physical justification, this explanation clearly does not work for simple or
doubly stable stratified fluids. On the other hand, exergy theory in thermodynamics offers a potentially more promising avenue. Exergy theory posits that 
because internal energy is convex with respect to its canonical variables
(specific entropy, salinity, and specific volume for two-component seawater), a portion of it --- referred to as exergy, proportional to the deviation from thermodynamic equilibrium --- is available for conversions into work 
\citep{Tailleux2013,Bannon2013,Bannon2014} under diabatic transformation, 
regardless of whether the 
stratifying tracers are stabilising or destabilising. For single-component fluids, exergy is directly 
proportional to temperature variance at leading order, providing a means to identify and quantify the portion of BPE that can be converted into APE in simple and doubly diffusive flows. However, this approach necessitates accepting APE dissipation as a conversion between APE and the internal energy component of BPE, 
thereby challenging the prevailing assumption. 
Pursuing this line of enquiry is beyond
the scope of this paper and is deferred to future studies. 
We anticipate that this will
require the development of a new class of Boussinesq models with explicit 
and thermodynamically consistent internal energy. The development of such a model,
building upon the recent findings of \citet{TailleuxDubos2024}, is currently 
under way and will be presented in a subsequent study. 
} 
\\
\smallskip \indent
In this study, the local APE framework was preferred over \citet{Winters1995} global APE framework \textcolor{black}{owing to its greater simplicity, physical justification, and flexibility}. Despite its popularity, the global APE framework has several unphysical aspects that can sometimes lead to erroneous conclusions and/or interpretations. Examples include its unsuitability for regional studies 
\textcolor{black}{in open domains due to the unphysical character of its advective fluxes}, and the inability to meaningfully decompose the sign-indefinite integrand $\rho g (z-z_r)$ into mean and eddy components. These issues, which tend to be overlooked in the turbulent mixing community, 
\textcolor{black}{can only be corrected by using the local APE framework}.
\textcolor{black}{While the non-intuitive character of the
early formulations of the local APE theory by 
\citet{Andrews1981}, \citet{Holliday1981}, and \citet{Shepherd1993} made it
somewhat difficult to use or apply, the most recent formulations have 
greatly clarified it and several applications have since demonstrated its usefulness
and ease of use, e.g., \citet{Roullet2009,Zemskova2015,Novak2018,Harris2018}.
Given that the local APE framework has made the global APE framework largely
obsolete, there is no longer any real justification for continuing using the
global framework, especially as it is obviously unsuitable for the study of 
individual turbulent mixing events in large domains such as that pertaining to the
oceans. We hope that the simplicity of our derivations, which we believe to be
more intuitive and transparent than those of \citet{Scotti2014}, will convince
some readers to switch frameworks. } 
\\
\smallskip \indent 
\textcolor{black}{For a single-component fluid, the Boussinesq APE dissipation rate 
$\varepsilon_p = {\bf J}_b \cdot \nabla (z-z_r)$ can be shown to approximate 
the following quantity
\begin{equation}
   \rho \varepsilon_{p} = \kappa c_p \nabla T \cdot 
   \nabla \left ( \frac{T-T_r}{T} \right ) 
   \label{exact_ape_dissipation}
\end{equation} 
}
\citep{Tailleux2009,Tailleux2013c}. 
Eq. (\ref{exact_ape_dissipation}) features a Carnot-like thermodynamic efficiency $(T-T_r)/T$ underscoring the inherent thermodynamic character of turbulent stratified mixing and APE dissipation, which is hidden by the Boussinesq approximation. \citet{Tailleux2013c} 
demonstrated that (\ref{exact_ape_dissipation}) could be broken down into three parts \textcolor{black}{(see Appendix A for a simpler and clearer derivation of the result)}: a laminar component $\varepsilon_{p,lam}$ related to the term $-\Phi_i$ in W95; a turbulent term $\varepsilon_{p,turb}$ related to the term $\Phi_d$ in W95; and a term $\varepsilon_{p,eos}$ that arises from the nonlinearities of the equation of state and has no equivalent in the standard Boussinesq model and W95.
\textcolor{black}{As far as we understand the issue, these components are 
indissociable parts of $\varepsilon_p$ and therefore necessarily of the same type,
thus implying that $\Phi_d$ and $\Phi_i$ should both be interpreted as 
conversions between APE and internal energy, in contrast to the prevailing
interpretation.}
In the literature, $\Phi_i$ or $\varepsilon_{p,lam}$ are often neglected based on the assumption that they are negligible in fully turbulent flows. However, this may not necessarily 
be the case, as $\Phi_i$ appears to be a significant term in many published studies. For example, $\Phi_i$ is comparable to $\Phi_d$ in the simulations discussed by W95.
\textcolor{black}{The present theory predicts that $\Phi_i$ should also be dominant in the early stages of the canonical salt finger instability experiment discussed by \citet{Middleton2020}; it is therefore unfortunate that the authors did not
diagnose $\Phi_i$ in their experiments, as this would undoubtedly have been useful
to demonstrate its fundamental importance.} 
We hope to illustrate this behaviour in the future by using an energy complete Boussinesq approximation currently under development and based on the findings of \citet{TailleuxDubos2024}. \\
\smallskip \indent

\textcolor{black}{Like \citet{Middleton2020}, we find that three physical parameters are sufficient to fully describe the behaviour of $\Phi_d$ or $\varepsilon_p$, but apart from the diffusivity ratio $\kappa_S/\kappa_T$, how to chose the two remaining parameters is not unique. The parameters chosen by \citet{Middleton2020}}
are strongly affected by turbulence and the micro-structures of $\theta$ and $S$, so diagnosing them from observational data or environmental parameters, or even linking them to parameters used in previous studies (such as the density ratio), is not straightforward. To overcome some of these difficulties, \citet{Middleton2021} subsequently reformulated the problem in terms of density/spiciness variable, using a linear spiciness variable to achieve a more physical and useful characterisation of $\Phi_d$. However, this spiciness variable is arguably suboptimal for measuring spiciness because it is not properly calibrated to vanish in a spiceless fluid.
\textcolor{black}{For this reason, \citet{Tailleux2021} recommended that spiciness be defined} as an isopycnal anomaly instead. \textcolor{black}{On this basis, we propsoed} a new set of physical parameters that we used to characterise $\varepsilon_p$: a density ratio and a new spiciness parameter more in line with the parameters used in previous studies. However, estimating these parameters in numerical simulations or in the field requires the computation of thickness-weighted isopycnal averages, which most non-oceanographers are likely to be unfamiliar with. 
For this reason and to facilitate the adoption of our ideas, we provided explicit mathematical expressions and practical examples to show how to implement this form of averaging in practice, valid even for a fully compressible and nonlinear ocean, and illustrated it on a realistic oceanographic example with real data. \\
\smallskip \indent

While the presence of double diffusive instabilities is determined by the sign of the full APE dissipation rate $\varepsilon_p = \varepsilon_{p,lam} + \varepsilon_{p,tur}$, the effective diapycnal diffusivity controlling the evolution of the sorted density profile is only determined by the turbulent part $\varepsilon_{p,tur}$. Consequently, the conventional expressions for turbulent diapycnal diffusivity and dissipation ratio should be formulated as $K_v = \varepsilon_{p,turb}/N^2$ and $\Gamma = \varepsilon_{p}/\varepsilon_k = \varepsilon_{p,turb}/\varepsilon_k$ rather than in terms of the full $\varepsilon_p$. This is consistent with the common practice, but it may not be necessarily obvious if one considers that $\varepsilon_{p,lam}$ (equivalently $\Phi_i$) is part of the definition of the net APE dissipation rate, which departs from the prevailing view that only $\varepsilon_{p,turb}$ or $\Phi_d$ contributes to it. As a consequence, not all double-diffusive instabilities can be expected to exhibit negative effective diffusivities, which, for instance, are not present in the initial stages of the canonical salt finger instability discussed by \citet{Middleton2020}. 
Our theory shows that double diffusion can enhance, reduce, or even change the sign of $K_v$, whose consequences for the study of mixing in the oceans need to be further investigated. The implications for the theoretical understanding of thermohaline staircases is unclear, as recent theories tend to emphasise the spatial variations (linked to the functional dependence on some turbulent parameters such as the buoyancy Reynolds number) of an otherwise positive effective diffusivity as the main causee, e.g., see \citet{Radko2005}, \citet{Taylor2017} or \citet{MaPeltier2022,MaPeltierJFM2022} and references therein for some recent discussion. Further research is clearly needed to clarify the issue, but beyond the scope of this paper. \\
\smallskip \indent 

\medskip 
\par \noindent \medskip
{\bf Declarations of interests}. The author reports no conflict of interest. 

\appendix 

\textcolor{black}{
\section{APE dissipation in compressible and Boussinesq fluids}
In contrast to the concept of viscous dissipation, whose origin can be traced back to the development of the Navier-Stokes equations in the early 19th century, the concept of APE dissipation is comparatively much more recent, as it only appears to have been developed in a somewhat ad-hoc way from a re-scaling of the dissipation of temperature variance by \citet{Gargett1984b} and \citet{Oakey1982}. In particular, it was not obtained as part of a local APE budget, even though the possibility to construct \citet{Lorenz1955} APE theory from a local principle had been established by \citet{Andrews1981} and \citet{Holliday1981}. Local balance equations for a local APE density was perhaps first derived by \citet{Molemaker2010}.
As far as we are aware, a theoretical formulation of APE dissipation for a fully compressible fluid clarifying how it relates to its Boussinesq counterpart was only developed much more recently by \citet{Tailleux2009} and \citet{Tailleux2013c}.
Here, we revisit our earlier 
derivation to provide a somewhat simpler treatment, in which dependence of our expressions in terms of entropy is replaced by a dependence on potential temperature.
Concretely, the task here is to 
understand how to relate the expression for the APE dissipation rate for a 
fully compressible fluid 
\begin{equation}
    \varepsilon_p = \kappa_T c_p \nabla T \cdot 
    \nabla \Upsilon = - \kappa_T c_p \nabla T \cdot
    \nabla \left ( \frac{T_r}{T} \right ) 
\end{equation}
to the terms $\Phi_i$ and $\Phi_d$ that appear in \citet{Winters1995} global APE
framework, where $T_r = T(\eta,p_r)$ and $T=T(\eta,p)$ 
are the in-situ temperature at the
reference pressure $p_r$ and actual pressure $p$ respectively,
with $\Upsilon = (T-T_r)/T$ the thermodynamic efficiency. 
}
\textcolor{black}{
As in \citet{Tailleux2013}, we take as our starting point the following 
thermodynamic relation for the total differential of specific entropy $\eta$
regarded either as a function of $(T,p)$ or $\theta$, viz., 
\begin{equation}
  {\rm d}\eta = \frac{c_p}{T} {\rm d}T 
  - \frac{\alpha}{\rho } {\rm d}p = \frac{c_{p\theta}}{\theta} {\rm d}\theta ,
  \label{thermo_relation_one}
\end{equation}
in which $\eta$ is the specific entropy, 
$\alpha$ is the thermal expansion coefficient, $c_p$ the specific heat
capacity at constant pressure, $\theta = T(\eta,p_a)$ the potential temperature,
i.e., the temperature reference to the mean surface atmospheric pressure $p_a$,
and $c_{p\theta} = c_p(\theta,p_a)$. The thermodynamic relation
(\ref{thermo_relation_one}) can be re-arranged to yield an expression for
the total differential of the logarithm of the temperature $\ln{T}$, viz., 
\begin{equation}
    \frac{{\rm d}T}{T} = \frac{\alpha}{\rho c_p} {\rm d}p 
+ \frac{c_{p\theta}}{c_p} \frac{{\rm d}\theta}{\theta} , 
\label{thermo_relation_two}
\end{equation}
thus allowing one to regard $T=T(\theta,p)$ as a function of potential 
temperature and pressure. Because ${\rm dT}/T$ is an exact differential,
Eq. (\ref{thermo_relation_two}) implies the
following Maxwell relationship (that is, the equality of the cross derivatives): 
\begin{equation}
     \frac{\partial}{\partial \theta} \left ( \frac{\alpha}{\rho c_p} \right ) 
     = \frac{c_{p\theta}}{\theta} 
     \frac{\partial}{\partial p} \left ( \frac{1}{c_p} \right ) 
     \label{maxwell_relationship} 
\end{equation}
which will prove useful in the following. 
For the cases of interest, the reference temperature entering calculations are `potential temperature' of some kind referenced to an arbitrary pressure $p_r$. 
It follows that if we denote $T_r = T(\theta,p_r)$, then we may write
\begin{equation}
   \ln{\frac{T_r}{T}} = \int_{p}^{p_r} \frac{\alpha}{\rho c_p}(\theta,\tilde{p})\,{\rm d}\tilde{p} 
\end{equation}
Taking the gradient, we then obtain
\begin{equation} 
   \frac{T}{T_r} \nabla \frac{T_r}{T} 
    = -\frac{\alpha}{\rho c_p} (\theta,p)\nabla p
    + \left [ \int_{p}^{p_r} \frac{\partial}{\partial \theta} 
    \left ( \frac{\alpha}{\rho c_p} \right ) (\theta,\tilde{p}) {\rm d}\tilde{p} 
    + \frac{\alpha}{\rho c_p} (\theta,p_r) 
    \frac{dp_r}{d\theta} \right ] \nabla \theta 
    \label{nabla_t_tr} 
\end{equation} 
Using the above Maxwell relationship (\ref{maxwell_relationship}) 
\begin{equation}
    \int_{p}^{p_r} \frac{\partial}{\partial \theta}
    \left ( \frac{\alpha}{\rho c_p} \right ) (\theta,\tilde{p}) \,{\rm d}\tilde{p} 
    = \frac{c_{p\theta}}{\theta} \int_{p}^{p_r}  
    \frac{\partial}{\partial p} \left ( \frac{1}{c_p}
    \right ) (\theta,\tilde{p}) \,{\rm d}\tilde{p}
    = \frac{c_{p\theta}}{\theta} \left ( \frac{1}{c_{pr}} - \frac{1}{c_p}
    \right ) 
\end{equation}
so that (\ref{nabla_t_tr}) may be rewritten as follows
\begin{equation}
    \nabla \left ( \frac{T_r}{T} \right ) 
    = - \frac{\alpha T_r}{\rho c_p T} \nabla p 
    + \frac{\alpha_r T_r}{\rho_R c_{pr} T} \nabla p_r 
    + \frac{c_{p\theta} T_r}{T\theta} 
    \left ( \frac{1}{c_{pr}} - \frac{1}{c_p} \right ) \nabla \theta 
\end{equation}
with $c_{pr} = c_p (\theta,p_r)$, $\alpha_r = \alpha(\theta,p_r)$, 
$\rho_r = \rho(\theta,p_r)$. Note here that the notation $Q=Q(\theta,p)$ is an abuse of notations for the function $Q=Q(T(\theta,p),p)$, i.e., it is computed from the expressions in terms of in-situ temperature and pressure in which $T$ is replaced by its expression as a function of potential temperature and pressure.  
}
\textcolor{black}{
The above results may be used to express the non-viscous nonconservation term $\varepsilon_A$ as the sum
\begin{equation}
      \varepsilon_p = \varepsilon_{p,lam} + 
      \varepsilon_{p,tur} + \varepsilon_{p,eos} 
\end{equation}
Moreover, if one use the fact that $p_r = p_0(z_r(\theta))$ so that
$\nabla p_r = -\rho_r g dz_r/d\theta \nabla \theta$, one may write 
}
\par \medskip 
\textcolor{black}{
\fbox{
\begin{minipage}{12cm} 
\begin{equation}
    \varepsilon_{p,lam} = \frac{\alpha T_r}{\rho T} \kappa_T \nabla T 
    \cdot \nabla p 
\end{equation}
\begin{equation}
    \varepsilon_{p,tur} =  \frac{c_p T_r \alpha_r g}{c_{pr}T}
    \frac{dz_r}{d\theta} \kappa_T \nabla T \cdot \nabla \theta  
\end{equation}
\begin{equation}
    \varepsilon_{p,eos} = - \frac{c_{p\theta}T_r}{c_{pr}T} 
    \frac{(c_p - c_{pr} )}{\theta}  
    \kappa_T \nabla T \cdot \nabla \theta 
\end{equation}
\end{minipage}
}
}
\par \medskip \noindent 
\textcolor{black}{
Importantly, the term $\varepsilon_{p,lam}$ vanishes only at standard thermodynamic
equilibrium, characterised by uniform in-situ temperature $T$. However, both 
$\varepsilon_{p,tur}$ and $\varepsilon_{p,eos}$ vanish both at standard thermodynamic equilibrium, but also for the `turbulent' thermodynamic equilibrium characterised by uniform potential temperature $\theta$. 
}

\textcolor{black}{
\subsection{Link with Boussinesq approximation} 
The above expressions show that the exact counterpart of the $\Phi_i$ term 
in \citet{Winters1995} is the following expression: 
\begin{equation}
    \Phi_i^{exact} = - \int_{V} \frac{\alpha T_r}{T} \kappa_T \nabla T 
    \cdot \nabla p \,{\rm d}V 
\end{equation}
\begin{equation}
   \Phi_d^{exact} = \int_{V} \frac{c_p T_r \rho \alpha_r g}{c_{pr} T}
   \frac{d z_r}{d\theta} \kappa_T \nabla T\cdot \nabla \theta \,  {\rm d}V 
\end{equation}
In the Boussinesq approximation, $p \approx -\rho_{\star} g z$, $T_r \approx T$,
and $\alpha$ is treated as a constant. As a result,
\begin{equation}
    \Phi_i^{exact} \approx  \int_{V} \rho_{\star} \alpha g 
    \frac{\partial T}{\partial z} \, {\rm d}V = 
    \rho_{\star} \alpha g \left ( \langle T \rangle_{top} - 
    \langle T \rangle_{bottom} \right ) = \Phi_i 
\end{equation}
As is well known, this term integrates to a term that depends on the difference
between the surface integrated temperature at the upper boundary minus that at
the lower boundary. For a single-component fluid, $\Phi_i$ is generally
positive for a stable stratification. In particular, $\Phi_i$ is independent of
the turbulent character of the fluid. It is important that this is not the case
of the exact expression. An important open question is whether strong turbulence
could potentially make $\Phi_i^{exact}$ significantly different from $\Phi_i$. 
}
\textcolor{black}{
Regarding the $\Phi_d$ term,
\begin{equation}
    \Phi_d^{exact} \approx \int_{V} \rho_{\star} \alpha g 
    \frac{dz_r}{d\theta} \kappa_T |\nabla \theta|^2 \,{\rm d}V 
    = - \int_{V} g \frac{dz_r}{d\rho} \kappa_T |\nabla \rho|^2 \,{\rm d}V 
\end{equation}
assuming ${\rm d}\rho = - \rho_{\star} \alpha {\rm d}\theta$. 
The Boussinesq approximation of $\Phi_d^{exact}$ is therefore potentially
much more accurate than that for $\Phi_i$. 
}
\textcolor{black}{
Note that 
\begin{equation}
    \varepsilon_{p,eos} \approx - \frac{c_{p} - c_{pr}}{\theta} 
    \kappa_T \nabla T \cdot \nabla \theta
\end{equation}
is related to a nonlinear equation of state. This term vanishes if the pressure
variations of $c_p$ can be neglected. As shown by the maxwell relation
\ref{maxwell_relationship}, this is the case if 
\begin{equation}
     \frac{\partial}{\partial \theta} \left ( 
     \frac{\alpha}{\rho c_p} \right ) = 0
\end{equation}
Therefore, the condition $\alpha/(\rho c_p) = {\rm constant}$ was taken as the
appropriate definition of a linear equation of state underlying the construction
of the Boussinesq approximation in \citet{Tailleux2013c}. 
}

\bibliographystyle{jfm}

\end{document}